\documentclass[twocolumn, times]{aastex63}
\usepackage[T1]{fontenc}
\usepackage{amsmath}
\usepackage{color}

\received{6/23/20}
\revised{6/10/21}
\accepted{9/9/21}

\submitjournal{ApJ}

\shorttitle{Discovery of Molecular Line Polarization in the Disk of TW~Hya}
\shortauthors{Teague et al.}

\graphicspath{{./}{figures/}{appendix_figures/}}

\newcommand{\todo}[1]{\textcolor{red}{[#1]}}
\newcommand{\tocite}[1]{\todo{citation needed}}

\begin{document}

\title{Discovery of Molecular Line Polarization in the Disk of TW~Hya}

\correspondingauthor{Richard Teague}
\email{richard.d.teague@cfa.harvard.edu}

\author[0000-0003-1534-5186]{Richard Teague}
\affil{Center for Astrophysics | Harvard \& Smithsonian,
       60 Garden Street, 
       Cambridge,
       MA 02138,
       USA}

\author[0000-0002-8975-7573]{Charles L. H. Hull}
\affil{National Astronomical Observatory of Japan,
       Alonso de C\'{o}rdova 3788,
       Office 61B,
       7630422,
       Vitacura,
       Santiago,
       Chile}
\affil{Joint ALMA Observatory,
       Alonso de C\'{o}rdova 3107,
       Vitacura,
       Santiago,
       Chile}
\affil{NAOJ Fellow}

\author[0000-0003-3773-1870]{St\'{e}phane Guilloteau}
\affil{Laboratoire d’Astrophysique de Bordeaux,
       Universit\'{e} de Bordeaux,
       CNRS, 
       B18N, 
       All\'{e}e Geoffroy Saint-Hilaire,
       33615 Pessac,
       France}

\author[0000-0003-4179-6394]{Edwin A. Bergin}
\affil{Department of Astronomy,
	   University of Michigan,
       311 West Hall,
       1085 S. University Ave,
       Ann Arbor,
       MI 48109,
       USA}
       
\author{Anne Dutrey}
\affil{Laboratoire d’Astrophysique de Bordeaux,
       Universit\'{e} de Bordeaux,
       CNRS, 
       B18N, 
       All\'{e}e Geoffroy Saint-Hilaire,
       33615 Pessac,
       France}

\author[0000-0002-1493-300X]{Thomas Henning}
\affil{Max Planck Institute for Astronomy,
      K\"{o}nigstuhl 17,
      69117 Heidelberg,
      Germany}
       
\author[0000-0003-2309-8963]{Rolf Kuiper}
\affil{Zentrum f\"{u}r Astronomie der Universit\"{a}t Heidelberg,
       Institut für Theoretische Astrophysik,
       Albert-Ueberle-Stra{\ss}e 2,
       69120 Heidelberg,
       Germany}

\author[0000-0002-3913-7114]{Dmitry Semenov}
\affil{Max Planck Institute for Astronomy,
       K\"{o}nigstuhl 17,
       69117 Heidelberg,
       Germany}
\affil{Department of Chemistry,
       Ludwig Maximilian University,
       Butenandtstr. 5-13,
       81377 Munich,
       Germany}

\author[0000-0003-3017-4418]{Ian W. Stephens}
\affil{Department of Earth, Environment, and Physics,
       Worcester State University,
       Worcester,
       MA 01602,
       USA}
\affil{Center for Astrophysics | Harvard \& Smithsonian,
       60 Garden Street,
       Cambridge,
       MA 02138,
       USA}
       
\author[0000-0002-2700-9916]{Wouter H. T. Vlemmings}
\affil{Department of Space, Earth and Environment,
       Chalmers University of Technology,
       Onsala Space Observatory,
       439 92 Onsala,
       Sweden}

\begin{abstract}
We report observations of polarized line and continuum emission from the disk of TW~Hya using the Atacama Large Millimeter/submillimeter Array. We target three emission lines, $^{12}$CO (3-2), $^{13}$CO (3-2) and CS (7-6), to search for linear polarization due to the Goldreich-Kylafis effect, while simultaneously tracing the continuum polarization morphology at 332\,GHz (900\,\micron{}), achieving a spatial resolution of 0.5\arcsec{} (30~au). We detect linear polarization in the dust continuum emission; the polarization position angles show an azimuthal morphology, and the median polarization fraction is $\sim$\,0.2\%, comparable to previous, lower frequency observations. Adopting a `shift-and-stack' technique to boost the sensitivity of the data, combined with a linear combination of the $Q$ and $U$ components to account for their azimuthal dependence, we detect weak linear polarization of $^{12}$CO and $^{13}$CO line emission at a $\sim 10\sigma$ and $\sim 5\sigma$ significance, respectively. The polarization was detected in the line wings, reaching a peak polarization fraction of $\sim 5\%$ and $\sim 3\%$ for the two molecules between disk radii of $0\farcs5$ and 1\arcsec{}. The sign of the polarization was found to flip from the blue-shifted side of the emission to the red-shifted side, suggesting a complex, asymmetric polarization morphology. Polarization is not robustly detected for the CS emission; however, a tentative signal, comparable in morphology to that found for the $^{12}$CO and $^{13}$CO emission, is found at a $\lesssim 3\sigma$ significance. We are able to reconstruct a polarization morphology, consistent with the azimuthally averaged profiles, under the assumption that this is also azimuthally symmetric, which can be compared with future higher-sensitivity observations.
\end{abstract}

\keywords{Circumstellar disks (235), Radio interferometry (1346), Polarimetry (1278), Submillimeter astronomy (1647)}

\section{Introduction}
\label{sec:introduction}

It is widely believed that magnetic fields play a key role in the star-formation process \citep{PlanckXXXV, Hull_Zhang_2019, PattleFissel2019}. During the collapse of a star-forming core, the coupling of the ionized gas to the ambient magnetic fields will dictate the rate of collapse and fragmentation of the material, influencing the formation of protoplanetary disks \citep[e.g.,][]{Wurster_Li_2018}. After a protoplanetary disks has formed, the presence of a magnetic field would readily account for the removal of angular momentum necessary to balance the observed accretion rates \citep[e.g.][]{Gammie_1996}.

Turbulent viscosity arising due to the magneto-rotational instability \citep[MRI;][]{Balbus_Hawley_1991} has long been regarded to be the most likely candidate for enabling angular momentum transport; however, recent searches for non-thermal broadening indicative of turbulent motions have failed to detect any significant motions in any disks except that around DM~Tau \citep{Hughes_ea_2011, Guilloteau_ea_2012, Flaherty_ea_2015, Teague_ea_2016, Flaherty_ea_2017, Teague_ea_2018b, Flaherty_ea_2020}. These results also suggest that other hydrodynamical instabilities, such as the vertical shear instability, may also be inactive given the lack of broadening observed \citep{Flock_ea_2017}. Given these results, magnetically driven winds \citep[e.g.,][]{Blandford_Payne_1982, Turner_ea_2014} appear to be a more promising route for removing angular momentum.

In order to test how magnetic fields can influence the formation and evolution of protoplanetary disks via the induction of such dynamical processes, observations tracing the magnetic field strength and morphology are required. Typically this is achieved through the observation of polarized emission, either through the linear or circular polarization. Polarized continuum emission from circumstellar disks is now routinely observed with the Atacama Large Millimeter/submillimeter Array \citep[ALMA; although detections have been made with the CARMA interferometer, e.g.,][]{Stephens_ea_2014}; the polarization patterns have a variety of morphologies, including azimuthal \citep{Kataoka2017, Harrison_ea_2019, Vlemmings_ea_2019} or a combination of azimuthal and radial \citep{Kataoka_ea_2016}; aligned with the minor axis of the disk \citep{Hull_ea_2018, Dent_ea_2019, Harrison_ea_2019}; some combination of both \citep{Stephens_ea_2014, Mori2019}; and patterns that change with wavelength, as in the case of HL Tau and DG Tau \citep{Stephens_ea_2017, Bacciotti2018, Harrison_ea_2019}.

Currently there are as many mechanisms to produce polarization as there are different emission morphologies. The most well known, and original motivation for studying polarized emission, is the magnetic alignment of grains via radiative alignment torques \citep[RATs;][]{LazarianHoang2007a, Andersson2015}, believed to be the dominant alignment mechanisms in star-forming cores \citep[e.g.,][]{Hull_Zhang_2019} and tentatively in the disk of HD~142527 \citep{Kataoka_ea_2016,Ohashi_ea_2018}. More recently, alignment of grains with respect to the radiation intensity gradient \citep{Tazaki_ea_2017, Yang_ea_2019} and mechanical alignment from the gas flow around the grains \citep{Gold_1952, Kataoka_ea_2019} have been proposed as potential causes for observed polarization patterns in disks. Finally, the inclusion of self-scattering of polarization emission has also been shown to reproduce a large sample of the observed emission morphologies \citep{Kataoka_ea_2015, Pohl_ea_2016, Yang_ea_2016}, and is likely to be the dominant source of continuum polarization at sub-millimeter wavelengths.

Given that several of the aforementioned mechanisms are able to produce polarized continuum emission without the need for a magnetic field, recent effort has focused on searching for polarized molecular line emission. \citet{Goldreich_Kylafis_1981, Goldreich_Kylafis_1982} showed that in the presence of a strong radiation field or velocity gradients, the magnetic sub-levels of atoms and molecules can be unevenly populated, giving rise to linear polarization; this is known as the Goldreich-Kylafis effect \citep[e.g.][]{Morris_ea_1985}. Emission from the Goldreich-Kylafis effect has been observed in a variety of sources, typically limited to molecular outflows in young stellar objects, and has been seen to have high polarization fractions, reach up levels of $\sim 10\%$ \citep[e.g.,][]{Girart1999, Beuther_ea_2010, Ching_ea_2016, Vlemmings_ea_2017, Lee_ea_2018}. Recently, \citet{Stephens_ea_2020} conducted a deep search for linear polarization arising from the disks around IM~Lup and HD~142527, finding tentative evidence for low-level polarization of $^{12}$CO (2-1) in both sources, with polarization fractions of $\sim 1\%$. In addition to linear polarization from the Goldreich-Kylafis effect, circular polarization has been used to trace the line-of-sight magnetic field strength via Zeeman-splitting of spectral line emission from molecular clouds \citep{Crutcher_2012}; however, in the case of protoplanetary disks, only two upper limits have been reported, in TW~Hya \citep{Vlemmings_ea_2019} and AS 209 \citep{Harrison_ea_2021}. Given the difficulties of disentangling the multiple possible origins of polarized continuum emission, searching for linearly polarized spectral-line emission appears to be the most robust approach for detecting and characterizing the magnetic field morphology and strength in a protoplanetary disk.
 
We present ALMA polarization observations of $^{12}$CO~(3-2), $^{13}$CO (3-2), and CS (7-6) molecular line emission and 332~GHz continuum emission in the disk around TW~Hya. We describe the observations, calibration and imaging in Section~\ref{sec:observations}. Next we describe an extensive search for polarized molecular line emission in Section~\ref{sec:searching}; in Section~\ref{sec:discussion} we discuss the detection of continuum polarization and the non-detection of  molecular-line polarization in the context of previous studies. We offer our conclusions in Section~\ref{sec:summary}.

\section{Observations}
\label{sec:observations}

\begin{deluxetable}{cccc}
\tabletypesize{\normalsize}
\tablecaption{Observation Log}
\label{tab:observations}
\tablehead{
\colhead{Date} &\colhead{Int. Time} & \colhead{\# Ant.} & \colhead{Mean PWV} \\ [-8pt]
\colhead{} &\colhead{(min)} & \colhead{} & \colhead{(mm)}
} 
\startdata
Dec 19, 2018 & 16.68 & 42 & 1.53 \\
Dec 19, 2018 & 29.10 & 43 & 1.42 \\ [5pt]
Dec 25, 2018 & 16.17 & 45 & 0.37 \\
Dec 25, 2018 & 23.82 & 45 & 0.47 \\ [5pt]
Apr  8, 2019 & 16.17 & 49 & 0.88 \\
Apr  8, 2019 & 11.72 & 47 & 0.96 \\ [5pt]
Apr  9, 2018 & 16.17 & 46 & 1.45 \\ [5pt]
Apr 10, 2019 & 15.17 & 43 & 1.08 \\
Apr 10, 2019 & 24.32 & 41 & 1.83 \\
\enddata
\end{deluxetable}

\begin{figure*}
    \centering
    \includegraphics[width=\textwidth]{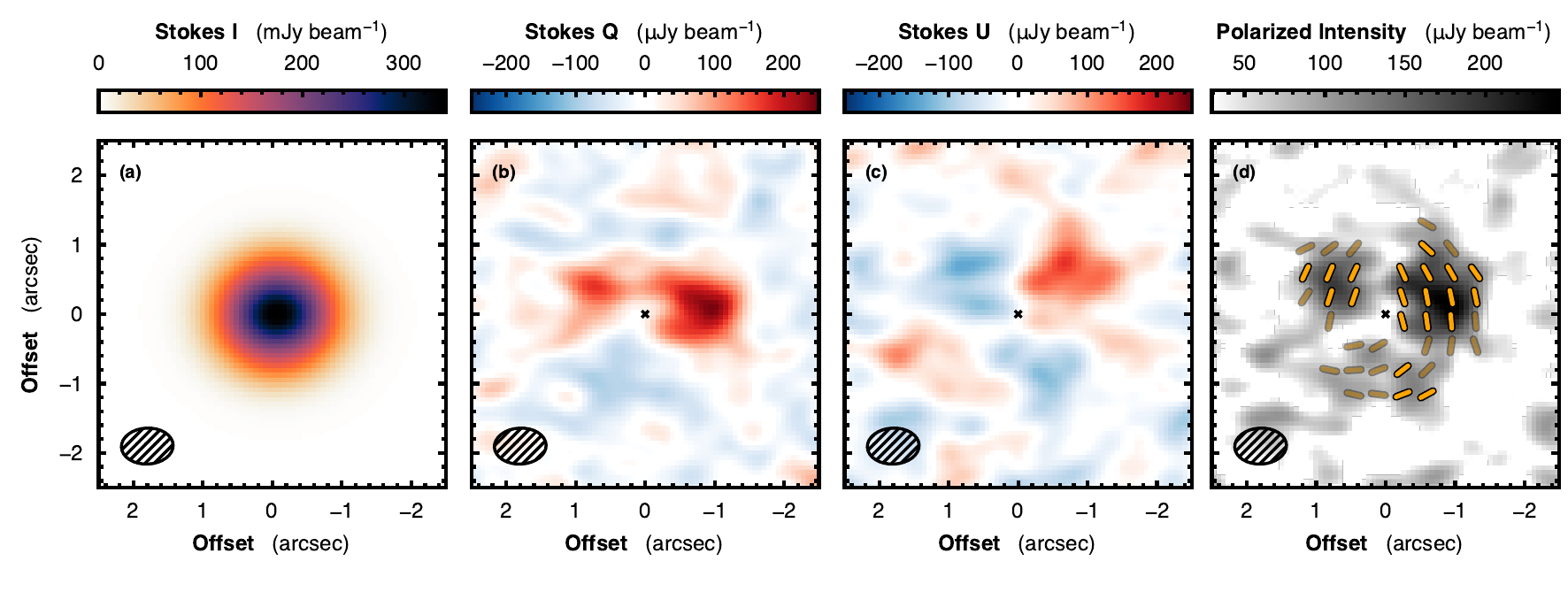}
    \caption{Summary of continuum observations. The Stokes $I$, $Q$, $U$ and polarized intensity $P$ images are shown from left to right, with the synthesized beam shown in the bottom left of each panel with a size of $0\farcs52 \times 0\farcs75$ and a position angle of $93\degr$. $P$ is shown at a significance of $>1\,\sigma_P$, where $\sigma_P = 32~\mu{\rm Jy\,beam^{-1}}$. In the right-most panel, the orange line segments show the polarization orientation. The semi-transparent segments show the polarization morphology in locations where $2\sigma_P < P < 3\sigma_P$, while all opaque bars indicate detections where $P>3\sigma_P$.}
    \label{fig:continuum_observations}
\end{figure*}

Observations were taken as part of project 2018.1.00980.S (PI: R. Teague), targeting $^{12}$CO (3--2), $^{13}$CO (3--2) and CS (7--6) in TW~Hya. The correlator was set up to include one continuum window centered at 331.75~GHz covering a bandwidth of 2~GHz, and three spectral windows centered on the three molecular lines with a spectral resolution of 61~kHz ($\approx 53~{\rm m\,s^{-1}}$). The C43-3 configuration was chosen to yield an angular resolution of $0.5\arcsec{}$. In total 9 execution blocks were observed, resulting in a total on-source time of 169.3 minutes. Table~\ref{tab:observations} provides a summary of the executions and the observing conditions. For all executions, the same phase (J1037-2934), flux (J1058+0133), and polarization calibrators (J1256-0547 and J1337-1257) were observed, and baselines spanned 14~m to 500~m.

\subsection{Calibration}
\label{sec:observations:calibration}

The data were first manually calibrated by NRAO/NAASC staff following standard ALMA procedures; for a detailed description of ALMA polarization calibration, see \citet{Nagai2016}.\footnote{Recently it was found that there is an error in the visibility amplitude calibration for sources with strong emission (\url{https://almascience.nao.ac.jp/news/amplitude-calibration-issue-affecting-some-alma-data}); however, there are currently no correction scripts available for polarization data. It is understood that this error should only affect amplitude calibration, and thus introduce an additional noise component to the polarization signal on the order of a few percent of the typical noise. As this work focuses on the extraction of a polarization signal through spectral shifting and stacking, it is unlikely that these amplitude calibration issues will influence the results presented here.}
We then self-calibrated the data using CASA v5.6.2 \citep{McMullin_ea_2007}. We used all continuum data, including the line-free channels of the three spectral windows covering the lines. We derived phase and amplitude self-calibration corrections based on the Stokes $I$ component of the continuum on a per-spectral-window basis (we did this, rather than collapsing all data into a single measurement, in order to avoid bandwidth smearing), and then applied these corrections to the entire data set. We performed three rounds of phase self-calibration (with solution times of `inf', 30\,s, and 10\,s) and one round of amplitude self-calibration. This improved the signal-to-noise ratio (SNR) of the Stokes $I$ continuum by a factor of 6.4 and achieved a noise in the Stokes $I$ images that is roughly three times the expected thermal noise.

\subsection{Continuum Emission}
\label{sec:observations:continuum}

We made 332~GHz (900~\micron{}) continuum images with a range of Briggs robust weighting values, spanning uniform (robust = --2, highest resolution and lowest sensitivity) to natural (robust = 2, lowest resolution and highest sensitivity) weighting. We detect linearly polarized Stokes $Q$ and $U$ emission when using a robust value of $\geq$\,0.5, without any significant change in the emission morphology as a function of robust parameter. As such, we adopt natural weighting (robust = 2) for the final image, which has a synthesized beam of $0\farcs75 \times 0\farcs52$ at a position angle of $93\degr$.

We produce $I$, $Q$, and $U$ images using the \texttt{tclean} function in CASA, adopting a circular mask that encompasses the full Stokes $I$ emission; this images are shown in Fig.~\ref{fig:continuum_observations}. For Stokes~$I$, the rms noise level measured in an emission-free region around the continuum emission in the non-primary-beam corrected image is $\sigma_{I} = 99~\mu{\rm Jy~beam}^{-1}$. The integrated continuum flux density is 1.38~Jy. For the $Q$ and $U$ components, we find rms noise values of $\sigma_Q = 33~\mu{\rm Jy~beam}^{-1}$ and $\sigma_U = 31~\mu{\rm Jy~beam}^{-1}$. Due to the technical challenges involved with calibrating circular polarization observations with ALMA, we focus here on the linear polarization observations; however, we do discuss the detection of circularly polarized (Stokes $V$) emission, which is most likely instrumental in nature, in Section~\ref{sec:discussion:circular_polarization}.

The polarized intensity $P$ and the polarization angle $\chi$ were then calculated as follows:

\begin{align}
    P    &= \sqrt{Q^2 + U^2}, \\
    \chi &= \frac{1}{2} \, {\rm tan}^{-1}\left(\frac{U}{Q}\right). \label{eq:polarization_angle}
\end{align}

\noindent As $P$ will always be positive, while both $Q$ and $U$ will be positive and negative, we debias $P$ following the procedure in \citet[but also see \citealt{Vaillancourt_2006, Hull_Plambeck_2015}]{Killeen_ea_1986}. Following this debiasing procedure,  described fully in Appendix~\ref{sec:app:debiasing}, we can assume that $\sigma_P \approx \sigma_Q \approx \sigma_U$ when $2 P \, / \, (\sigma_{\rm Q} + \sigma_{\rm U}) \gtrsim 1$.

We show a map of $P$, where $P \, / \, \sigma_P > 1$, in the right-hand panel of Fig.~\ref{fig:continuum_observations}. The line segments show the linear polarization orientation $\chi$. Segments are plotted at the Nyquist rate, i.e., every half-beam full-width at half-maximum (FWHM), in locations where $P \, / \, \sigma_P > 3$. The median continuum polarization fraction is 0.20\%, with a peak of 0.74\%. The observed azimuthal morphology of the polarization and the polarization fraction are comparable to 226~GHz continuum observations published in \citet{Vlemmings_ea_2019}, which we discuss in more detail in Section~\ref{sec:discussion:continuum}.

\begin{deluxetable*}{rcccccc}
\tabletypesize{\small}
\tablecaption{Imaging Summary\label{tab:images}}
\label{tab:lines}
\tablehead{
    \colhead{Line} &
    \colhead{Frequency} &
    \colhead{Channel Width} &
    \colhead{Beam (PA)} &
    \colhead{Integrated Flux Density\tablenotemark{\footnotesize \,a}} &
    \colhead{$\sigma_{I}$\tablenotemark{\footnotesize \,b}} &
    \colhead{$\sigma_{QU}$\tablenotemark{\footnotesize \,b,c}} \\
    \colhead{} &
    \colhead{(GHz)} &
    \colhead{(${\rm m\,s^{-1}}$)} &
    \colhead{} &
    \colhead{(${\rm Jy~beam^{-1}~km~s^{-1}}$)} &
    \colhead{(${\rm mJy~beam^{-1}}$)} &
    \colhead{(${\rm mJy~beam^{-1}}$)}
}
\startdata
$^{12}$CO (3--2) & 345.7959899 & 30 & $0\farcs60 \times 0\farcs44 \,\, (92\fdg3)$ & $44.91 \pm 0.01$    & 4.4 & 4.4 \\ [5pt]
$^{13}$CO (3--2) & 330.5879652 & 30 & $0\farcs66 \times 0\farcs47 \,\, (92\fdg6)$ & $\phn5.94 \pm 0.01$ & 5.0 & 5.0 \\ [5pt]
CS (7--6)        & 342.8828503 & 30 & $0\farcs60 \times 0\farcs44 \,\, (92\fdg3)$ & $\phn2.40 \pm 0.04$ & 3.4 & 3.4 
\enddata
\tablenotetext{a}{Integrated out to radii of $4\arcsec{}$ and over $2.84 \pm 2~{\rm km\,s^{-1}}$.}
\tablenotetext{b}{Measured in a line free channel.}
\tablenotetext{c}{Assuming both Stokes components share the same noise, verified in Section~\ref{sec:searching:CDFs}.}
\end{deluxetable*}

\subsection{Line Emission}
\label{sec:observations:line_emission}

\begin{figure*}
    \centering
    \includegraphics[width=\textwidth]{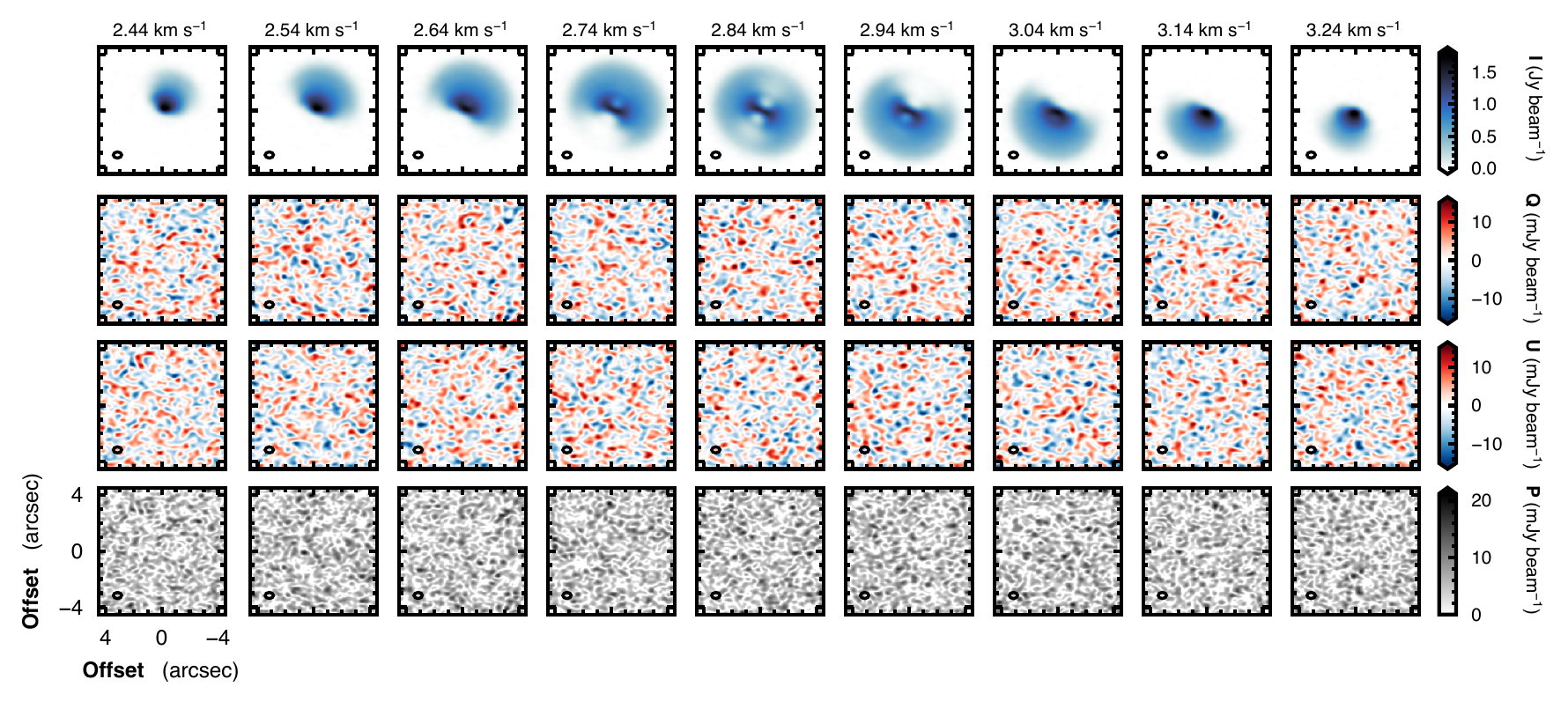}
    \caption{Channel maps of the $^{12}$CO (3--2) emission imaged at 100~${\rm m\,s^{-1}}$ intervals. The top row shows the total Stokes $I$ intensity, while the middle rows show the Stokes $Q$ and $U$ components, and the bottom row shows the debiased linear polarized intensity $P$. We only show the channels close to the systemic velocity, $v_{\rm LSR} = 2.84~{\rm km\,s^{-1}}$. We plot the beam size, with properties given in Table~\ref{tab:lines}, in the bottom-left corner of each panel. All images have been corrected for the primary beam.}
    \label{fig:12CO_channels}
\end{figure*}

\begin{figure*}
    \centering
    \includegraphics[width=\textwidth]{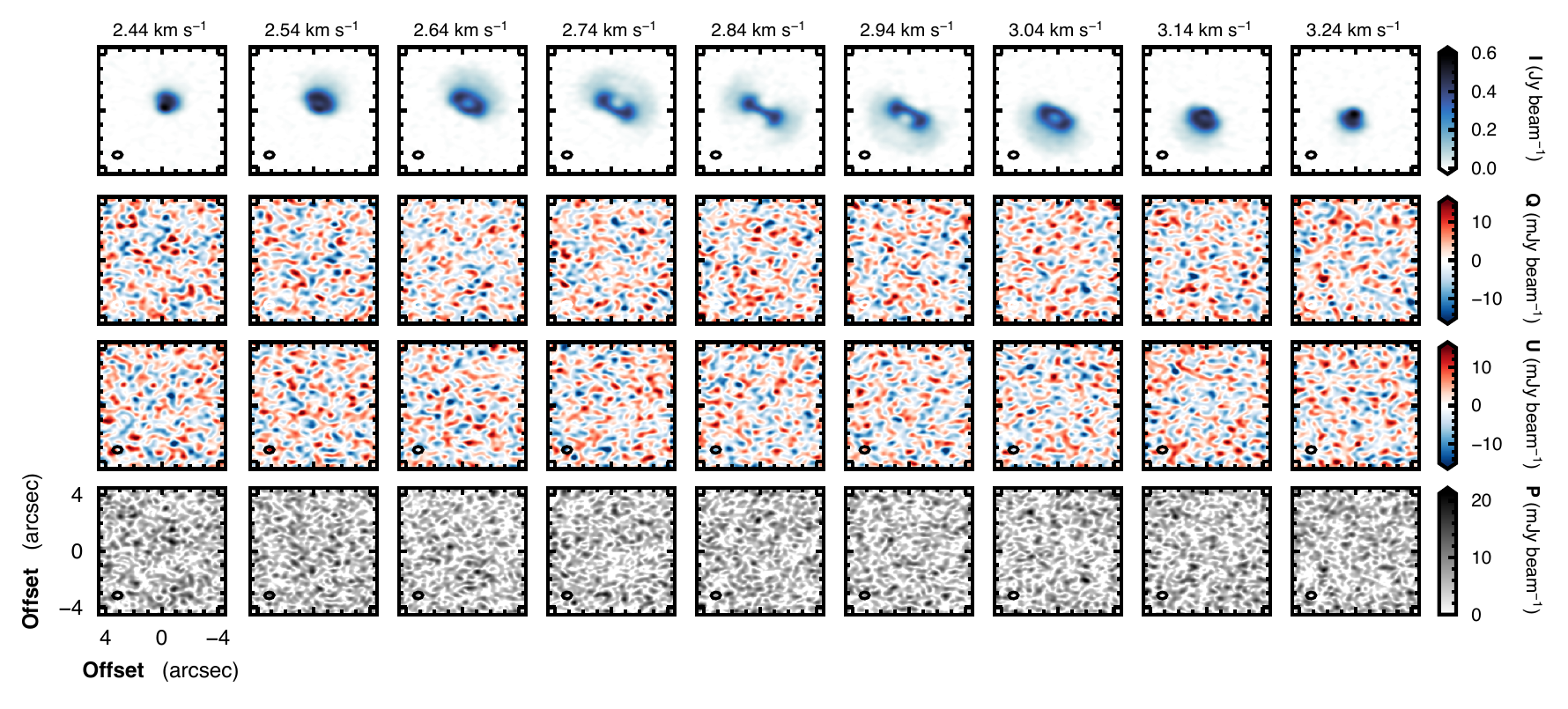}
    \caption{As Fig.~\ref{fig:12CO_channels} but for $^{13}$CO (3--2). }
    \label{fig:13CO_channels}
\end{figure*}

\begin{figure*}
    \centering
    \includegraphics[width=\textwidth]{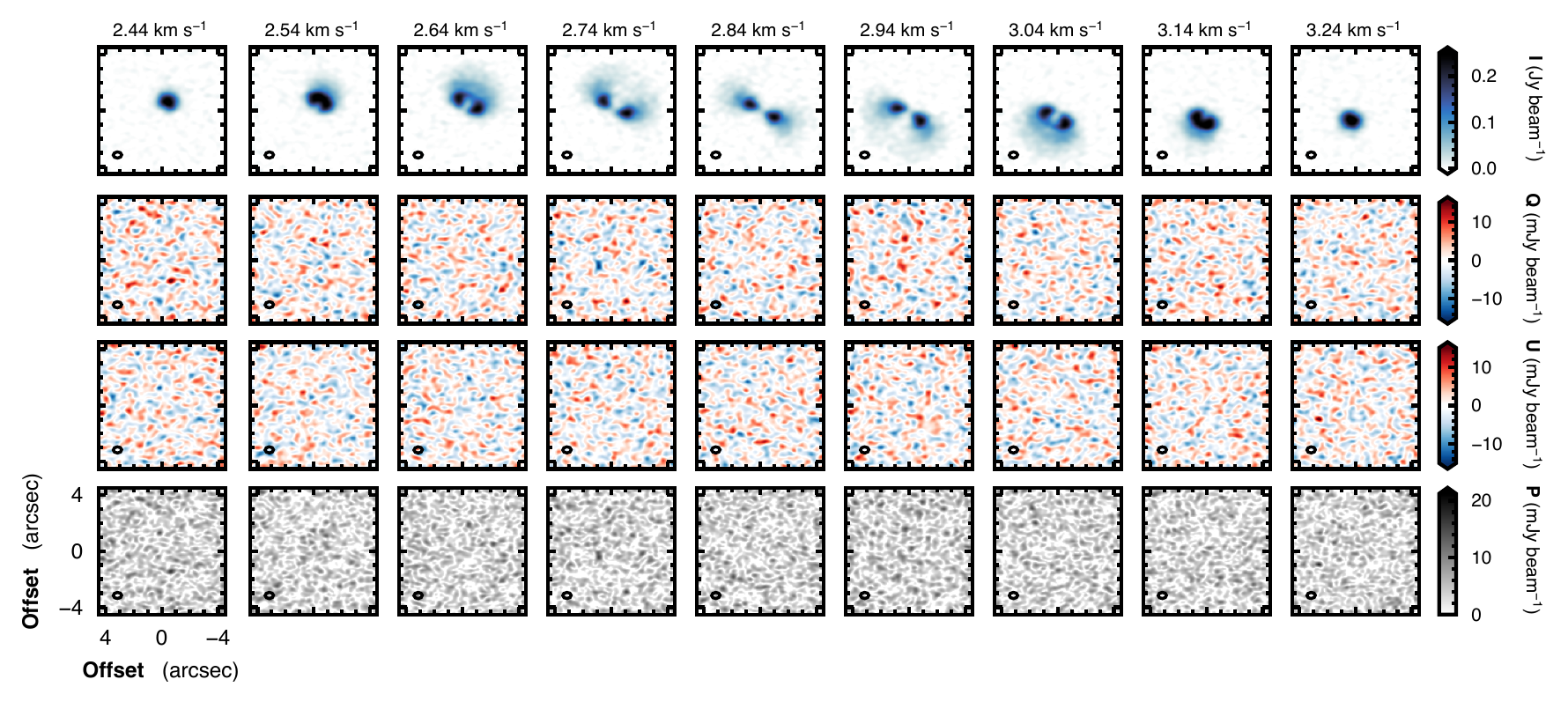}
    \caption{As Fig.~\ref{fig:12CO_channels} but for CS (7--6).}
    \label{fig:CS_channels}
\end{figure*}

Continuum emission was subtracted from the line emission using the CASA task \texttt{uvcontsub} which uses a linear fit to the line-free channels to model the continuum emission. This is expected to have little impact on the molecular emission, except in the inner regions of the disk where the continuum opacities are large and may lead to an underestimation of the peak brightness temperature \citep[see][for a discussion]{Boehler_ea_2017}.

We image the lines using a channel width of $30~{\rm m\,s^{-1}}$. This is close enough to Nyquist sampling to minimize artifacts when using the shifting and averaging techniques described below, while preserving any channel-to-channel correlations associated with the spectral response function.

As with the continuum, we made images at using range of robust values, spanning uniform weighting to natural weighting. We do not detect any obvious polarization in any maps, thus we adopt a robust value of 0.5 for the final image, resulting in a resolution of $\sim 0.6\arcsec$, as summarized in Table~\ref{tab:lines}. The choice of a lower robust value than for the continuum is because our line-averaging technique benefits from a larger number of independent samples, which are increased with higher angular resolution. As the radial extent of the continuum is $\sim 1\arcsec{}$, adopting a lower robust value would not significantly improve the number of independent measurements in the disk, but would decrease our sensitivity.

\begin{figure*}
    \centering
    \includegraphics[width=\textwidth]{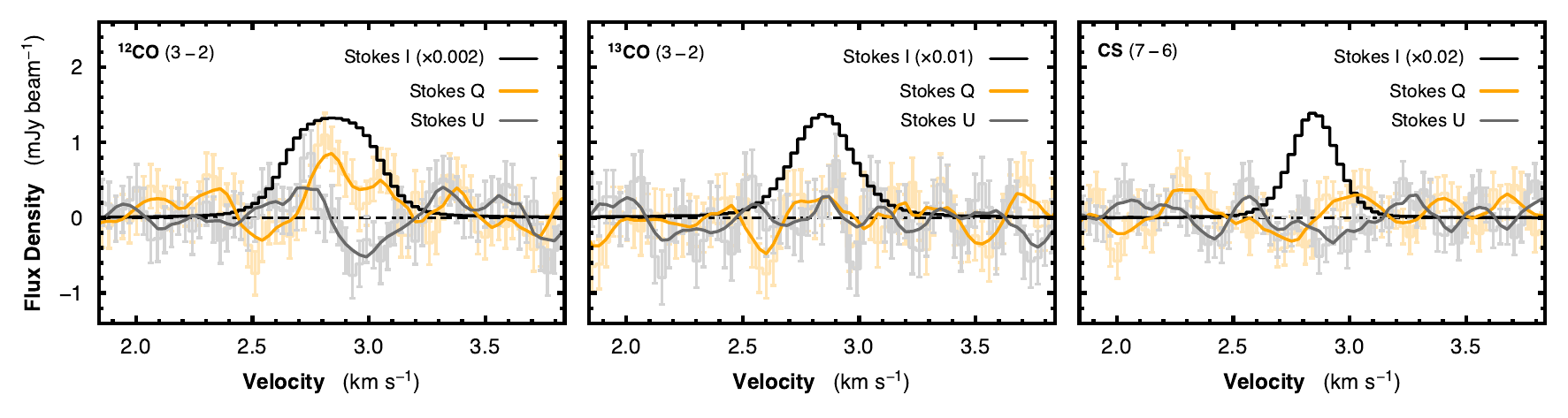}\vspace{0.5cm}
    \includegraphics[width=\textwidth]{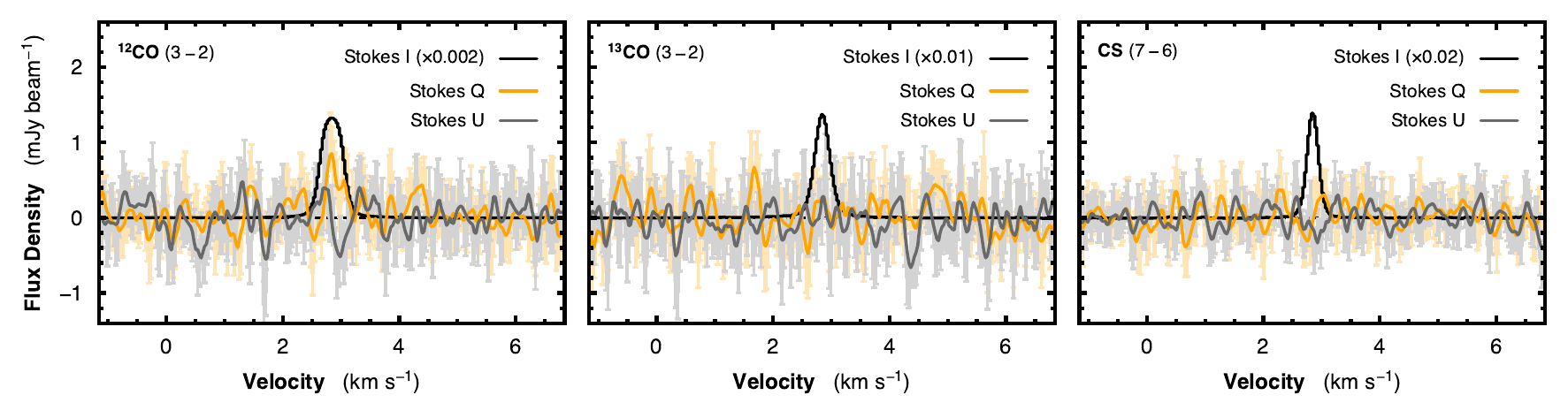}
    \caption{\emph{Top:} The disk-averaged spectra after aligning and averaging all components. $^{12}$CO, $^{13}$CO, and CS emission are shown from left to right.  The black, orange, and gray lines show the Stokes $I$, $Q$, and $U$ components, respectively. The Stokes $I$ components have been scaled by the factor shown in the legend in order to be plotted on the same $y$-axis. The semi-transparent lines show the aligned and averaged spectra at a channel spacing of $50~{\rm m\,s^{-1}}$, with error bars representing their associated $1\sigma$ uncertainties. The solid orange and gray lines show the average spectrum smoothed with a top-hat function with a $250~{\rm m\,s^{-1}}$ width, in order to bring out any underlying trends in the spectrum. Note that unlike the $^{13}$CO and CS emission, the $^{12}$CO Stokes $I$ component deviates from a Gaussian profile, as the line is highly optically thick. \emph{Bottom:} As the top row, but with extended velocity ranges.}
    \label{fig:averaged_spectra}
\end{figure*}

Prior to making the images, for each line we made a dirty image of five consecutive line-free channels, which we used to measure the noise in each of the Stokes components by taking the rms of the non-primary-beam-corrected values. We use this noise level as the threshold for a non-interactive \texttt{tclean} using a Keplerian mask\footnote{The code to make Keplerian masks can be found at \url{https://github.com/richteague/keplerian_mask}.} tailored to the $^{12}$CO emission, which also provides good fits to the $^{13}$CO and CS emission. We set the \texttt{tclean} stopping threshold to twice the noise level measured in the line-free channels. We present a summary of the images in Table~\ref{tab:images}; in Figures~\ref{fig:12CO_channels}, \ref{fig:13CO_channels} and \ref{fig:CS_channels} we show the channel maps (binned down to $100~{\rm m\,s^{-1}}$ channel spacing for presentation purposes) of the Stokes $I$, $Q$, and $U$ components, and of the debiased polarization maps $P$.

We calculate integrated flux densities of the Stokes $I$ components using \texttt{GoFish} \citep{GoFish}; see Section~\ref{sec:searching:stacking} for a more thorough description of this method. We adopt the disk properties reported in \citet{Teague_ea_2019a}: inclination angle $i = 5.8\degr$; disk position angle ${\rm PA} = 151.6\degr$, measured to the red-shifted major axis from North; stellar mass $M_{\rm star} = 0.81~M_{\rm sun}$; and distance $d = 60.1$~pc \citep{Bailer-Jones_ea_2018}. We note that there has been a claim of a slight warp in the inner 20~au of the disk \citep{Rosenfeld_ea_2012}; however, given the limited spatial resolution of these data ($\approx$\,30~au), the influence of the warp can be ignored. The recovered $^{12}$CO and CS integrated intensities are within 10\% of previously reported observations \citep{Huang_ea_2018, Teague_ea_2018b}, consistent with the typical flux calibration uncertainty for Band 7 observations with ALMA, as discussed in the ALMA Technical Handbook \citep{ALMATechnicalHandbook}.

\section{Searching for Spectral Line Polarization}
\label{sec:searching}

In contrast to the continuum images, the channel maps of the spectral lines show no clear sign of polarized emission, either in the native $Q$ and $U$ maps or in the debiased polarized intensity $P$ maps. In this section we apply three methods to attempt to place tighter limits on the presence of any polarization signals in the data.

\subsection{Azimuthal Averaging by Line Shifting}
\label{sec:searching:stacking}

In the context of protoplanetary disks, the use of azimuthally averaging line data by correcting for the Doppler shift of the disk rotation is becoming more common. Originally described in \citet{Yen_ea_2016}, and used in several contemporaneous works \citep[e.g.][]{Teague_ea_2016, Matra_ea_2017}, this approach has been used to detect weak spectral-line emission that is not clearly visible in standard interferometric channel maps \citep[e.g.][]{Schwarz_ea_2019}, and has been applied to the searches for spectral-line polarization signals in disks \citep[e.g.,][]{Stephens_ea_2020, Harrison_ea_2021}.

This approach assumes that the line-of-sight velocity $v_0$ is the sum of the systemic velocity $v_{\rm LSR}$ and the projected orbital rotation:

\begin{equation}
    v_0 = v_{\phi} \cos(\phi) \sin(i) + v_{\rm LSR}\,\,,
\end{equation}

\noindent where $v_{\phi}$ is the rotational velocity, $\phi$ is the deprojected polar angle such that $\phi = 0$ corresponds to the red-shifted major axis of the disk, and $i$ is the disk inclination. Assuming that $v_{\phi}$ is dominated by Keplerian rotation, it is possible to calculate the projected velocity component at every pixel in an image, allowing for spectra to be shifted back to a common line center and then averaged, thereby increasing the SNR of the line. In this work we use the Python package \texttt{GoFish}\footnote{\bf \url{https://github.com/richteague/gofish/releases/tag/v1.4.1-1}} \citep{GoFish}, which implements this aligning and averaging for arbitrary disk geometries and source properties.

Using this azimuthal averaging approach, we calculate the average spectrum between radii of $0\farcs{}5$ and $4\farcs{}5$ for the $^{12}$CO emission, and between $0\farcs{}5$ and $3\farcs{}5$ for the $^{13}$CO and CS emission. We avoid the inner $0\farcs{}5$ region of the disk as the line averaging method is significantly biased due to the strong spatial correlations on spatial scales comparable to or smaller than the beam FWHM \citep[e.g.,][]{Teague_ea_2018a}. In this calculation the disk is first split into annuli with widths of $0\farcs{}1$, using the same disk properties as those used to measure the integrated flux density in Section~\ref{sec:observations}. The spectra are then aligned by taking into account the projected Keplerian rotation of the disk, and finally are averaged at each radius. An average spectrum is calculated for each of the $I$, $Q$, and $U$ components. To estimate the uncertainty in these spectra, we image four additional cubes of line-free data at $\pm 9$ and $\pm 18~{\rm km~s^{-1}}$ offsets relative to the systemic velocity $v_{\rm LSR} = 2.84~{\rm km~s^{-1}}$. From these line-free cubes we produce averaged spectra from which we measure the rms noise, which we assume to be constant across the spectrum. When measuring the rms noise, we use velocity offset rather than a spatial offset because the point source sensitivity of the observations drops precipitously for off-axis positions due to the response of the primary beam.

We show the resulting averaged spectra in Fig.~\ref{fig:averaged_spectra}. In each panel, the Stokes $I$ component, shown in black, has been scaled down by a factor of 500, 100, and 50 for $^{12}$CO, $^{13}$CO, and CS, respectively. Error bars on these components are too small to be seen in the figure. The $Q$ and $U$ components, shown in orange and gray, respectively, are binned to their native spectral resolution of $30~{\rm m\,s^{-1}}$; we plot their associated $1\sigma$ uncertainties. To highlight any trends in the data, we plot a smoothed spectrum (solid line) using a top-hat function with a $250~{\rm m\,s^{-1}}$ width. No signal in either the $Q$ or $U$ components can be seen for either the $^{13}$CO or CS emission. The $^{12}$CO emission shows a tentative feature in both the $Q$ and $U$ components, with a positive peak in $Q$ at $v \approx 2.8~{\rm km~s^{-1}}$ and a negative peak in $U$ at $v\approx 2.9~{\rm km\,s^{-1}}$. However, features of a similar magnitude are seen across the whole velocity range (shown in the bottom row of Fig.~\ref{fig:averaged_spectra}), precluding the confirmation of a true signal.

\subsection{Teardrop Plots}
\label{sec:searching:teardrop}

\begin{figure*}
    \centering
    \includegraphics{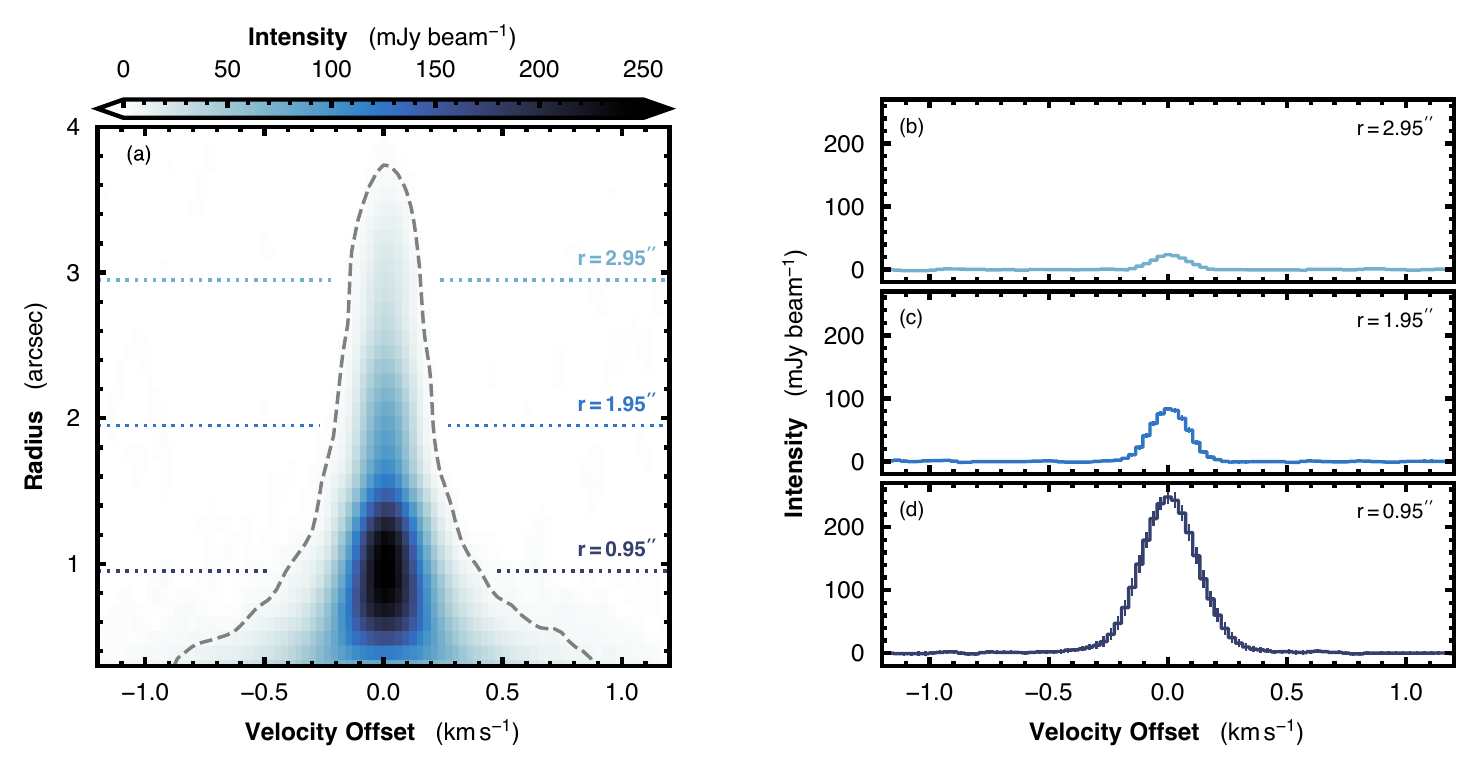}
    \caption{Example of a ``teardrop'' plot made with \texttt{GoFish} using the CS (7--6) spectra. The left panel, (a), shows the aligned and averaged spectra across the radius of the disk. Each row in this panel represents the average spectrum at that radius. The gray dashed line denotes $3\sigma$, where $\sigma$ is the radially varying noise (radially varying because at larger radii there are more beams to average over, thus reducing the noise). The right panels show cuts at constant $r$ (see the dotted lines in panel (a)). Error bars are not included, as they are small enough that they are not visible. Note that CS emission peaks at $r \approx 1\arcsec{}$, hence the offset peak in panel (a), which would otherwise lie at $r = 0\arcsec$.}
    \label{fig:teardrop_example}
\end{figure*}

An extension to the aligning and averaging approach is to use a modified position-velocity diagram (which we call a ``teardrop'' plot), which is a radially resolved counterpart to the disk-averaged spectrum discussed in the previous section. As demonstrated in Fig.~\ref{fig:teardrop_example}, each row in a teardrop plot represents the aligned and averaged spectrum at a certain radius in the disk. As the spectra have all been aligned due to the velocity shift, all line centers align along the systemic velocity of $2.84~{\rm km\,s^{-1}}$. At larger radii, the line becomes weaker, but also narrower due to the reduced thermal broadening from the lower temperature gas. The radially decreasing intensity can be easily seen by taking horizontal cuts at various radii, as shown in the right hand column of Fig.~\ref{fig:teardrop_example}, while the change in the width is most noticeable for disk radii $\lesssim 1\arcsec$. A slightly more complex aspect of this representation of the data is that the rms noise, $\sigma$, estimated using line-free regions, is no longer constant, but rather dependent on the radius. This is because at larger radii, more independent samples are used in the averaging such that $\sigma(r) \propto r^{-0.5}$. Note that this only describes the thermal noise; the radially varying sensitivity due to the primary beam response will result in radially increasing $\sigma$ in the primary-beam corrected images, such that $\sigma$ will drop off at a slightly slow rate than $r^{-0.5}$.

\begin{figure*}
    \centering
    \includegraphics[width=\textwidth]{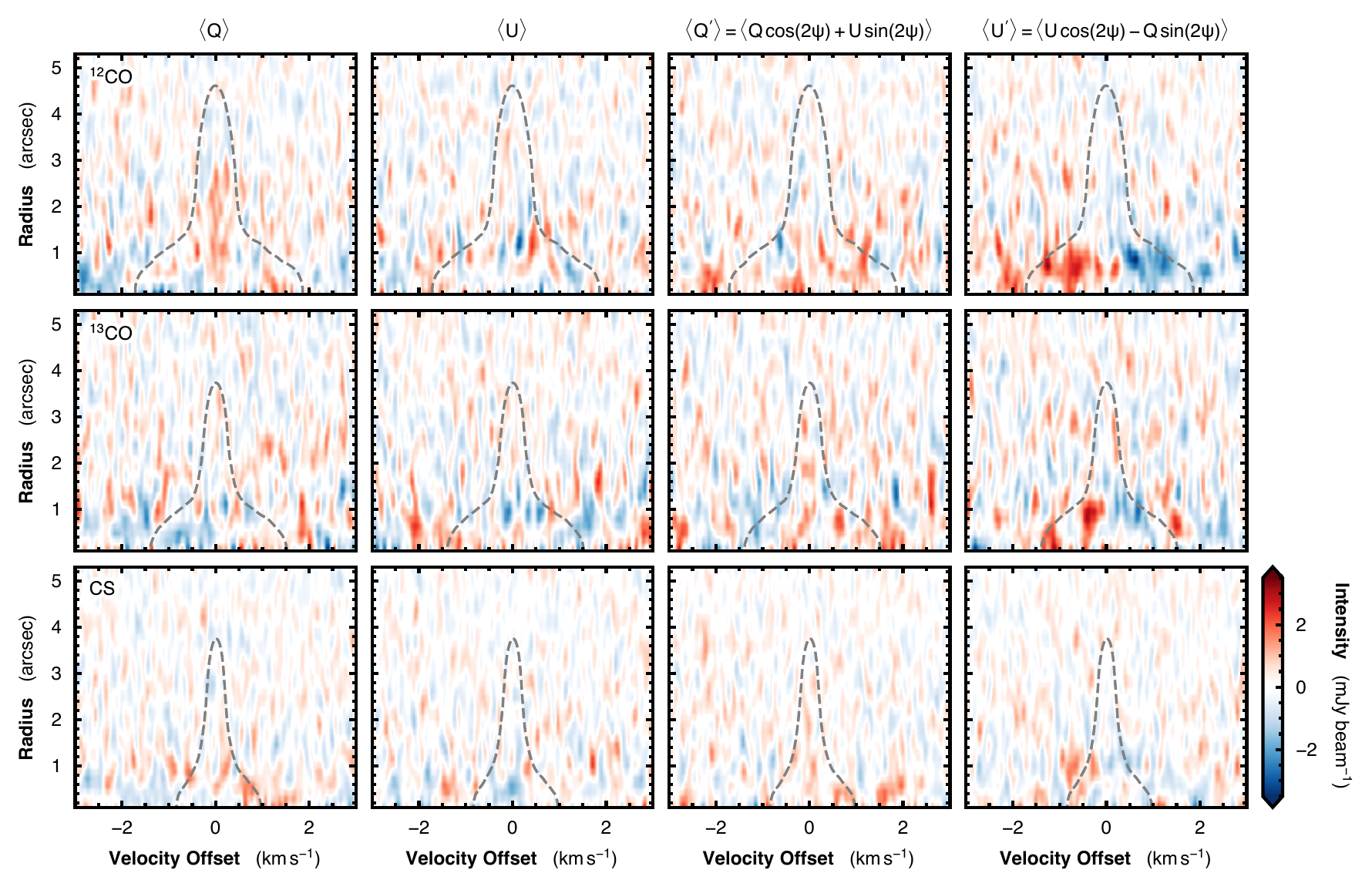}
    \caption{Teardrop plots for $^{12}$CO, $^{13}$CO, and CS emission (top to bottom) showing the $Q$, $U$, $Q^{\prime}$, and $U^{\prime}$ components (left to right). Each plot has the same color scaling. The dashed gray line shows the (radially varying) $3\sigma$ contour of the $I$ component to highlight where the spectral-line emission should lie. Polarization signals are seen in the $U^{\prime}$ column for both $^{12}$CO and $^{13}$CO, manifesting as positive and negative signals in the line wings at radii of $r \lesssim 1\arcsec{}$.}
    \label{fig:QUteadrops}
\end{figure*}

We first consider a simple azimuthal average, as in Section~\ref{sec:searching:stacking}, with the results shown in the left two columns of Fig.~\ref{fig:QUteadrops}, where the gray dashed outline shows the $3\sigma$ contour of the Stokes I component. No clear signal is seen in either the Stokes $Q$ or $U$ component for any of the three molecules, as expected from the disk-averaged approach described in the previous section.

However, for an azimuthally symmetric polarization morphology, the $Q$ and $U$ components have a strong azimuthal dependence. For example, for the case of a radial polarization morphology given by $(Q_{\rm rad},\, U_{\rm rad})$, or a circular polarization morphology given by $(Q_{\rm circ},\, U_{\rm circ})$, we find,
\begin{align}
    (Q_{\rm rad},\, U_{\rm rad}) &\propto (\cos(2\psi),\, \sin(2\psi))\,\,,\\
    (Q_{\rm circ},\, U_{\rm circ}) &\propto (\cos(2\psi + \pi),\, \sin(2\psi + \pi))\,\,,
\end{align}
where $\psi$ is the position angle, measured East of North. As such, an average of any of these components over the interval $(-\pi,\, +\pi)$ will average out to zero. To account for this azimuthal dependence, we define two linear combinations of these components that should allow for a non-zero signal when azimuthally averaged. We define,
\begin{align}
    Q^{\prime} = Q \cos(2\psi) + U \sin(2\psi)\,\,,
    \label{eq:transformsI}\\
    U^{\prime} = U \cos(2\psi) - Q \sin(2\psi)\,\,,
    \label{eq:transformsII}
\end{align}
such that for a radial polarization morphology, $(\langle Q^{\prime}\rangle, \langle U^{\prime}\rangle) = (+1,\, 0)$, where $\langle Q^{\prime} \rangle$ denotes the average over the full $2\pi$ azimuth, and for a circular polarization morphology, $(\langle Q^{\prime}\rangle, \langle U^{\prime}\rangle) = (-1,\, 0)$. We discuss in more detail this combination and the expected values given different polarization morphologies in Section~\ref{sec:discussion:lines}.

The results for these two combinations are shown in the third and fourth columns of Fig.~\ref{fig:QUteadrops} for each of the three molecules. No clear signal is seen for the $Q^{\prime}$ component, but both the $^{12}$CO and $^{13}$CO emission appear to show a $\langle U^{\prime} \rangle$ signal at the $\sim 3~{\rm mJy~beam^{-1}}$ level in the line wings for $r \lesssim 1\arcsec{}$, corresponding to a polarization fraction of $P_{\rm frac} \sim 5\%$ and $\sim 3\%$ for $^{12}$CO and $^{13}$CO, respectively. Integrating over the blue- and red-shifted wings of the line ($\pm 1.5~{\rm km\,s^{-1}}$ for $^{12}$CO and $\pm 0.5~{\rm km\,s^{-1}}$ for $^{13}$CO), these polarization signals are detected at the $\sim 10\sigma$ and $\sim 5\sigma$ level for each line wing of $^{12}$CO and $^{13}$CO, respectively. The CS emission tentatively shows a similar $U^{\prime}$ morphology, but at a much lower level and would require deeper observations for confirmation. For the CS emission, we find peak $\langle U^{\prime} \rangle$ values of $\sim 2~{\rm mJy~beam^{-1}}$, relating to a $\lesssim 3\sigma$ significance when integrating over $\pm 0.5~{\rm km\,s^{-1}}$. We note that the peak $\langle U^{\prime} \rangle$ values are smaller than the rms noise for a given channel, as detailed in Table~\ref{tab:lines}, and thus the application of this `shift-and-stack' approach is imperative to tease out these signals.

As we find $Q^{\prime} = 0$ and $U^{\prime} \neq 0$ for all cases, it is possible to rule out purely circular or purely radial polarization morphologies, such as those predicted for several models of well ordered magnetic fields \citep[e.g.,][]{Lankhaar_Vlemmings_2020, Lankhaar_ea_2021}. The interpretation of this signal is discussed in Section~\ref{sec:discussion:lines}.

\subsection{Cumulative Distribution Functions}
\label{sec:searching:CDFs}

\begin{figure*}
    \centering
    \includegraphics[]{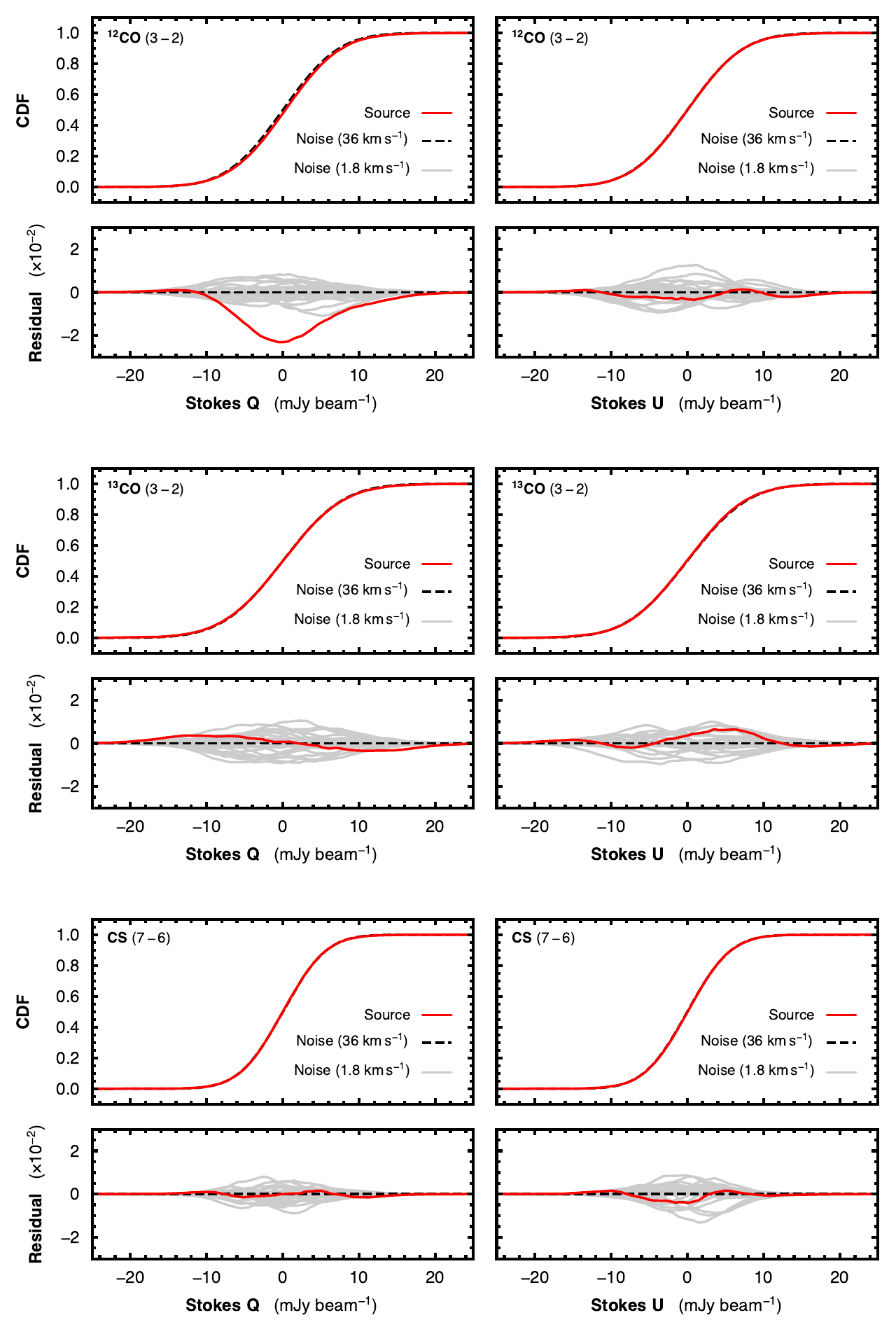}
    \caption{Cumulative distribution functions (CDF) of the Stokes $Q$ and $U$ components for all three molecular lines: $^{12}$CO (top), $^{13}$CO (middle), and CS (bottom). The top window of each panel shows the CDF of the flux density within an aperture centered on TW~Hya within a radius of 3\arcsec{}. The red line shows a $1.8~{\rm km\,s^{-1}}$ window centered on the systemic velocity ($v_{\rm LSR}$) of $2.84~{\rm km\,s^{-1}}$ which covers the spatially resolved Stokes $I$ emission. An additional 20 bands of $1.8~{\rm km\,s^{-1}}$, all line free, are shown in gray (although they cannot be seen behind the red line in the top panels). The bottom window of each panel shows the residual difference between the CDF in red and the average CDF made from all of the line-free regions in gray. The dashed black line shows zero difference.}
    \label{fig:CDFs}
\end{figure*}

Without priors on what the emission distribution for $Q$ and $U$ should look like, it is hard to distinguish peaks in the noise from true, low level signals. In this subsection we aim to characterize the properties of the noise in order to better understand whether the features discussed in the previous section are statistically significant.

We can compare the cumulative distribution functions (CDF) of the $Q$ and $U$ intensities with those obtained from line-free data. For this we use the four additional image cubes produced at velocity offsets of $\pm9$ and $\pm18~{\rm km\,s^{-1}}$ discussed in Section~\ref{sec:searching:stacking}. These four cubes are used to make a reference noise CDF. To account for the primary beam, we used images that have not been corrected for the primary beam; however, we focus on the inner 3\arcsec{} radius of the image, equivalent to roughly the inner third of the primary beam, which has a FWHM of $\sim 15.5\arcsec$ at the wavelength of our observations. The noise for both $Q$ and $U$ is well described by a Gaussian centered at zero with a standard deviation comparable to the rms values quoted in Table~\ref{tab:images}.

We extract the signal over a $1.8~{\rm km\,s^{-1}}$ window centered at the systemic velocity of $2.84~{\rm km\,s^{-1}}$ and that is broad enough to encompass all molecular emission.  We take only values within a 3\arcsec{} radius of the phase center, shown in red in Fig.~\ref{fig:CDFs}. To quantify the noise, we calculate additional CDFs using a similar $1.8~{\rm km\,s^{-1}}$ bandwidth from the line-free data to quantify the expected scatter due to smaller number statistics, shown as gray lines in Fig.~\ref{fig:CDFs}. The bottom window of each panel set shows the residual with respect to the average CDF made from all of the line-free regions, for the source in red, and the noise cubes in gray. All components except the $^{12}$CO Stokes $Q$ show deviations that are consistent with random draws from the line-free data (illustrated by the scatter in the gray lines). However, the $^{12}$CO Stokes $Q$ component does show a non-negligible deviation, with a peak residual of $\approx -2.3 \times 10^{-2}$. The deviation in $^{12}$CO Stokes $Q$ appears to be somewhat larger than the scatter found for just the noise values. The negative residual signals a slightly higher mean flux density value, which leads to a larger sample of positive flux densities; this is consistent with the tentative peak found in Section~\ref{sec:searching:stacking}, and suggests the presence of a tentative weak, but positively valued, Stokes~$Q$ signal, consistent with the $\langle U^{\prime} \rangle$ detection described above. The lack of clear signal in these statistical demonstrates the importance of leveraging prior knowledge of the polarization morphology in order to extract the signal from within the noise.

\section{Discussion}
\label{sec:discussion}

\subsection{Dust continuum Polarization}
\label{sec:discussion:continuum}

\begin{figure}
    \centering
    \includegraphics[width=\columnwidth]{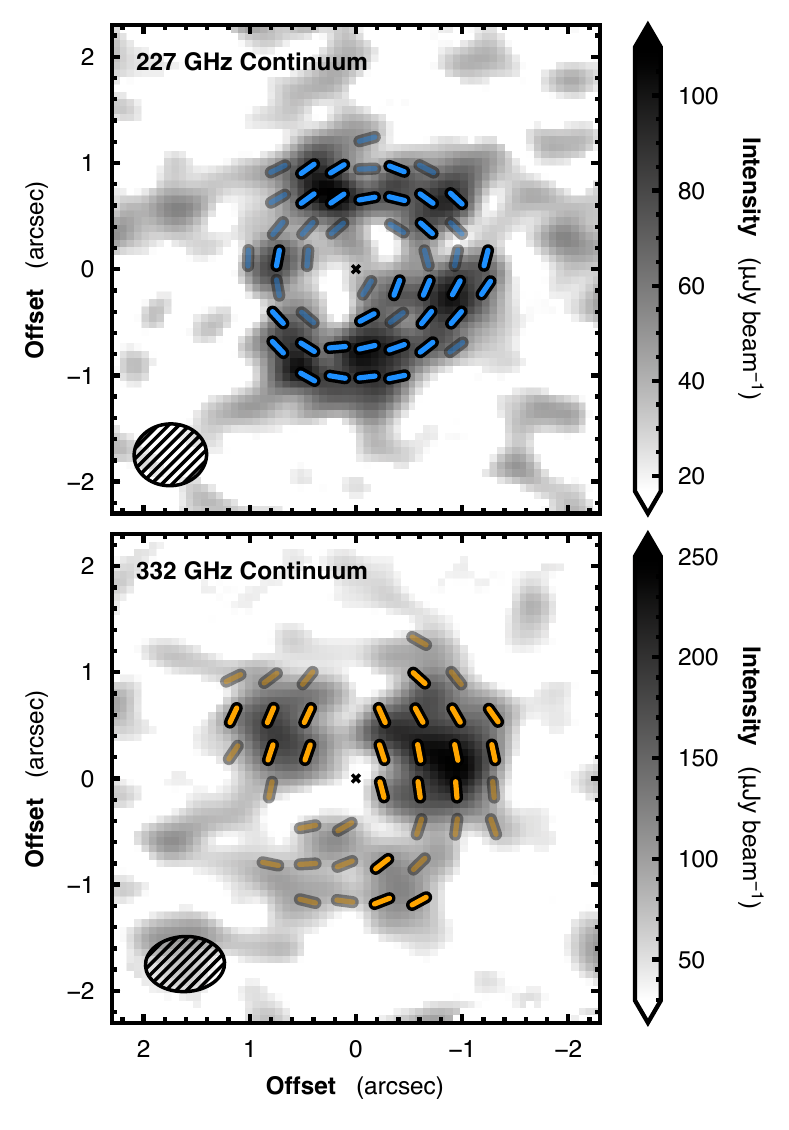}
    \caption{Comparison of 227~GHz continuum linear polarization from \citet{Vlemmings_ea_2019}, (\emph{left}), with the 332~GHz continuum from these observations (\emph{right}). The gray scale is plotted where $P > \sigma_P$, where $\sigma_P = 20$ and $35~\mu{\rm Jy~beam^{-1}}$ for the 227~GHz and 332~GHz data, respectively. The semi-transparent segments show the polarization morphology in locations where $2\sigma_P < P < 3\sigma_P$, while all opaque bars indicate detections where $P>3\sigma_P$. Segments are plotted at the Nyquist rate, i.e., every half-beam FWHM. The beam sizes are shown in the bottom-left corner of each panel.}
    \label{fig:continuum_comparison}
\end{figure}

Despite the lack of high significance spectral-line polarization, we robustly detect linear polarization in the continuum image, as shown in Fig.~\ref{fig:continuum_observations}. We compare this polarization morphology with the 227~GHz Band 6 continuum observations presented in \citet{Vlemmings_ea_2019} in Fig.~\ref{fig:continuum_comparison}. We re-imaged these latter data using the self-calibrated data products described in \citet{Vlemmings_ea_2019}, using natural weighting to yield a beam size of $0\farcs68 \times 0\farcs58$ ($93\degr$), comparable to that of our 332~GHz Band 7 continuum.

Both polarization morphologies are similar, showing polarization orientations that are approximately azimuthal. The median $P_{\rm frac}$ for the 227~GHz continuum observations is 0.32\%, while the 332~GHz continuum data has a median $P_{\rm frac}$ of 0.19\%. For both frequencies, $P_{\rm frac}$ decreases with radius, reaching peaks of $\sim 0.65\%$; however, this radial gradient in $P_{\rm frac}$ is most likely due to the large intensity gradient of Stokes $I$, which is poorly resolved at this spatial resolution. We note that this lack of spatial resolution is likely the limiting factor in the interpretation of the polarization morphology, particularly as TW~Hya is known to host extensive substructure in its continuum emission \citep{Andrews_ea_2016} which previous modeling efforts have shown can imprint non-trivial substructure in the polarization morphology \citep[e.g.,][]{Kataoka_ea_2015, Pohl_ea_2016}.

\begin{figure*}
    \centering
    \includegraphics{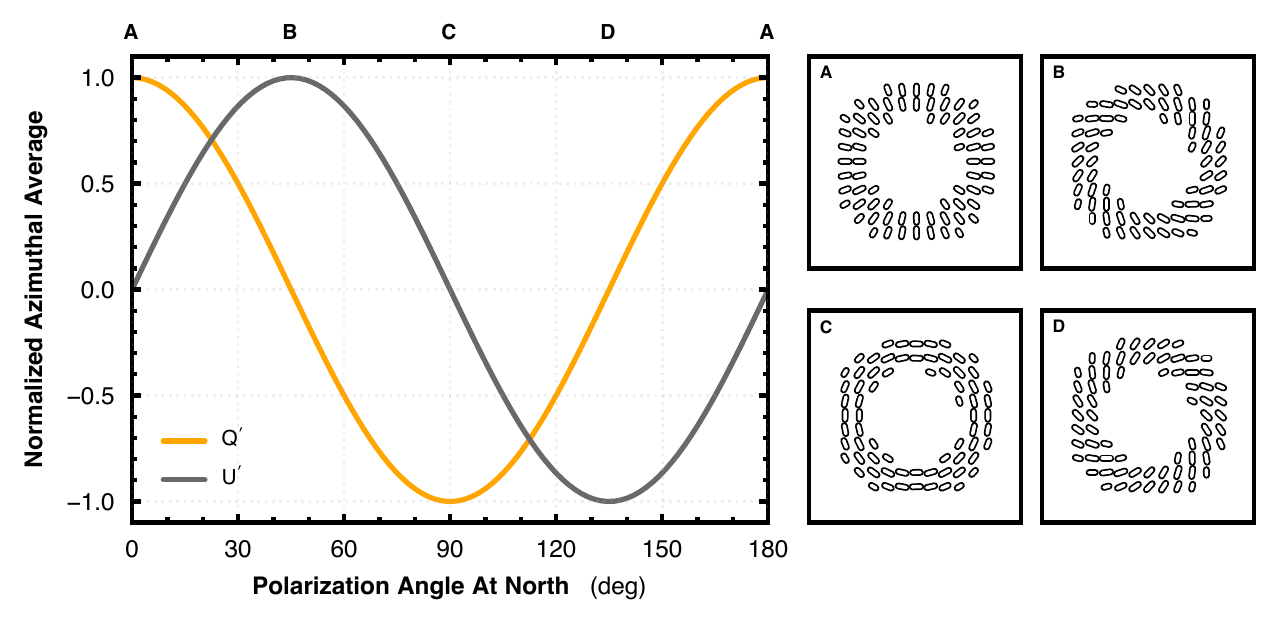}
    \caption{Demonstrating the azimuthally averaged value of the transformed $Q^{\prime}$ and $U^{\prime}$ components, defined in Equations~\ref{eq:transformsI} and \ref{eq:transformsII}, for a range of azimuthally symmetric polarization morphologies. The four panels on the right represent four possible scenarios, annotated in the left hand panel.}
    \label{fig:azimuthal_dependence}
\end{figure*}

The detection of such azimuthal morphologies in a face-on disk unfortunately do not allow us to distinguish easily among the various mechanisms that can produce dust continuum polarization. In the case of magnetically aligned dust grains, the polarization should be perpendicular to the magnetic field lines, na\"{i}vely suggesting a radial magnetic field structure, which is physically unlikely in a protoplanetary disk where magnetic field lines are likely to be toroidally wrapped by Keplerian rotation \citep[e.g.,][]{Flock_ea_2015, Suriano_ea_2018}. However, \citet{Guillet_ea_2020} showed that when dust grains have grown to a size similar to the observation wavelength (i.e., one is no longer in the Rayleigh regime but in the Mie regime), polarization from magnetically aligned dust grains can sometimes be \textit{parallel} to the magnetic field, potentially reconciling this mechanism with the broadly azimuthal morphology observed in TW~Hya. Despite this potential agreement, it is highly unlikely that the grains are magnetically aligned given the overwhelming evidence for alternative polarization mechanisms in other sources.

Self-scattering is particularly en vogue for protoplanetary disks, as scattering models are easily able to match observed polarization morphologies \citep[for example, but not limited to,][]{Kataoka_ea_2016, Hull_ea_2018, Dent_ea_2019, Harrison_ea_2019}. For the face-on case of TW~Hya, self-scattering could readily explain the azimuthal morphology, as we are tracing the edge of the continuum with our spatial resolution \citep[see, for example, models featured in][]{Kataoka_ea_2015, Pohl_ea_2016, Dent_ea_2019}. In addition to self-scattering, both alignment of dust grains with respect to the radiation intensity gradient (sometimes called ``radiative alignment''; see \citealt{Tazaki_ea_2017}) and mechanical alignment could be responsible for the observed emission morphology \citep{Yang_ea_2019, Kataoka_ea_2019}.

Several works have leveraged the polarization fraction as a probe of the maximum grain size within a disk, assuming that self-scattering is the dominant polarization mechanism. This is a particularly powerful way to infer changes in the grain size distribution across radial features such as gaps or rings \citep[e.g.,][]{Pohl_ea_2016, Dent_ea_2019}. Assuming that self-scattering dominates the polarization pattern observed in TW Hya, and adopting the aforementioned average polarization fractions of 0.35\% and 0.19\% for the Band 6 (1.3~mm) and Band 7 (0.9~mm) observations, we find a maximum grain size of $\gtrsim~350~\micron$ when comparing to Fig.~3 from \citep{Kataoka_ea_2016}. We note that this is an order of magnitude smaller than the maximum grain sizes inferred in \citet{Macias_ea_2021} who used a multi-wavelength analysis to infer the properties of the emitting grains. However, the large uncertainties in the average polarization fractions, the small spectral lever arm between Band 6 and Band 7 observations, and the difference in the angular resolution of the observations between these two methods (a factor of ${\sim}~20$) preclude a more detailed comparison.

Future multi-frequency observations and modeling efforts, such as those presented in \citet{Stephens_ea_2017}, may help to better constrain the underlying polarization mechanisms at play. However, given the relatively low spatial resolution of the polarization observations toward TW Hya, we defer such work to the future when higher spatial resolution observations will allow us to characterize in detail the polarization morphology of the dust continuum emission.

\subsection{Spectral-Line Polarization}
\label{sec:discussion:lines}

In Section~\ref{sec:searching}, we showed evidence for low-level linearly polarized emission from both $^{12}$CO and $^{13}$CO in the inner regions ($r \lesssim 1\arcsec{}$) of the disk via a `teardrop' deprojection of the data. This evidence is consistent with the CDF of pixel values found in the Stokes $Q$ cube for $^{12}$CO (see Fig.~\ref{fig:CDFs}), which suggest that there is at least some linearly polarized signal present.

\subsubsection{Polarization Fraction}

In Section \ref{sec:searching} we place tight limits on the polarization fraction of the $^{12}$CO, $^{13}$CO, and CS emission. Although we detect no polarized emission in the channel maps, we find a low-level signal when combining the $Q$ and $U$ components in a way that accounts for their azimuthal dependence. We find $\langle U^{\prime} \rangle \sim \pm 3~{\rm mJy~beam^{-1}}$ in the line wings of $^{12}$CO at radii $\lesssim 1\arcsec{}$, and tentative signs of a similar polarization signal in $^{13}$CO, also at radii $\lesssim 1\arcsec{}$. Assuming that the polarization is shared equally between the $Q$ and $U$ components, this suggests that the peak $Q$ and $U$ values are $\sim \sqrt{2}\,\langle U^{\prime} \rangle_{\rm max} \sim 4~{\rm mJy~beam^{-1}}$, comparable to the sensitivity of the observations (Table~\ref{tab:lines}), demonstrating the power of the `shift-and-stack' approach in leveraging knowledge of the velocity structure of the disk to tease out weak signals. Given that the polarization is detected in the line wings, this corresponds to a polarization fraction of $\sim 1\%$. No significant polarization is detected for CS emission. \citet{Hull2020b} note that an instrumental effect, known as 'squash', can lead to uncertainties of $\sim 0.5\%$ for both $Q$ and $U$ components, however such systematic effects are unlikely to be the cause of the polarization detected here as they should not have any spectral dependence, unlike the signal detected here.

At first glance, the low levels of polarization are surprising given the moderate spectral-line polarization fractions detected in star-forming sources \citep[][]{Beuther_ea_2010, Ching_ea_2016, Lee_ea_2018}; however, the local physical conditions for a dense protoplanetary disk differ substantially from these outflow regions. This fact will impact the level of observed polarization. Recent models of linear molecular-line polarization using the radiative transfer code \texttt{PORTAL} \citep{Lankhaar_Vlemmings_2020, Lankhaar_ea_2021} have shown that the level of polarization due to the Goldreich-Kylafis effect is highly dependent on the local physical conditions of where the polarized emission is emitted. The differential populations of the magnetic sub-levels are dictated by the relative rates of radiative and collisional interactions between molecules, with an enhanced rate of radiative interactions resulting in higher fractional polarization. Thus, in regimes where the rotational $J$-transitions are thermalised, collisional interactions will dominate and any polarization is likely to be suppressed. This can potentially account for the lack of strong polarization observed in TW~Hya, as all three molecular lines are believed to be fully thermalised \citep{Schwarz_ea_2016, Teague_ea_2018b}, unlike in the more tenuous outflow regions where the Goldreich-Kylafis effect has previously been detected.

\begin{figure*}
    \centering
    \includegraphics[width=\textwidth]{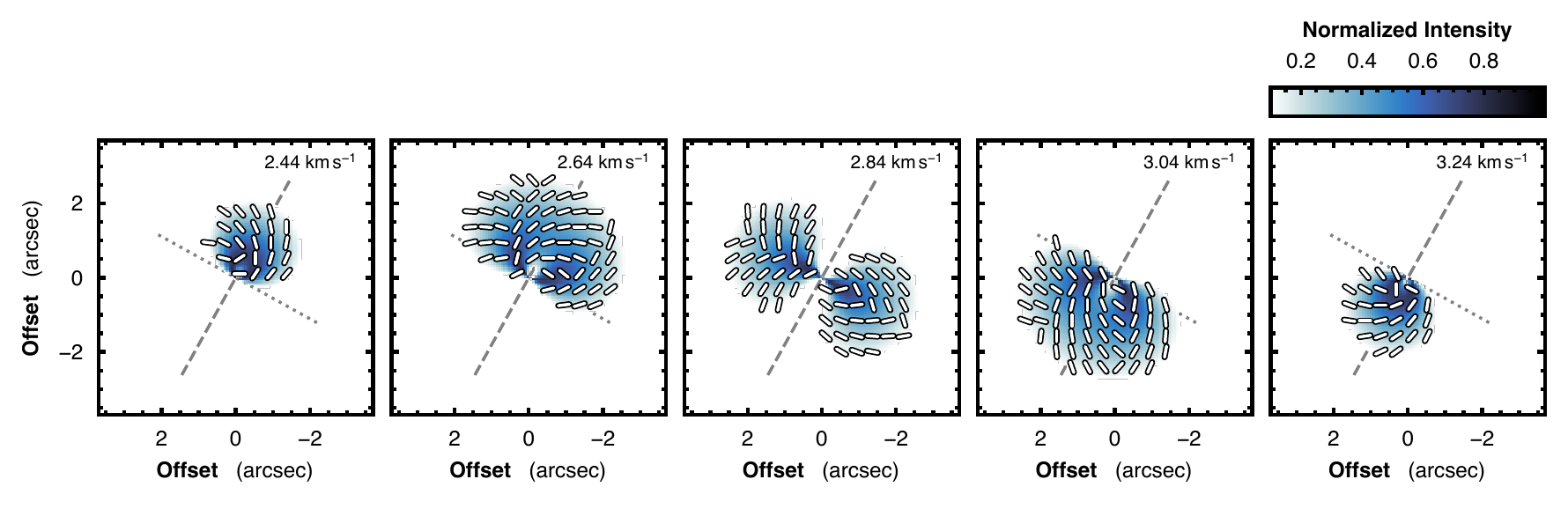}
    \caption{A proposed polarization morphology, using a simple 2D model of the line emission, that is consistent with the signal detected in the $\langle Q^{\prime} \rangle$ and $\langle U^{\prime} \rangle$ teardrop plots. To better demonstrate the morphology, we do not scale the lengths of the polarization line segments with the corresponding polarization fraction. The dashed gray lines show the major and minor axes of the disk, dashed and dotted, respectively. The velocity of each channel is shown in the top right of each panel.}
    \label{fig:model_polarization}
\end{figure*}

Due to the large densities expected in protoplanetary disks, it is unlikely that there are many, if any, lines which are both sub-thermally excited, such that the magnetic sub-levels are populated, and bright enough that the polarized emission could be detected directly. The best candidate would be HCN (4-3), which has both a high critical density and (faint) satellite hyperfine components which would provide a direct test of the optical depth. Based on previous detections of HCN \citep[e.g.][]{Guzman_ea_2015}, such observations would require integrations on the order of 10~hours in order to detect the predicted polarization fractions in TW~Hya.

\subsubsection{Morphology}

Using the transformations described by Equations~\ref{eq:transformsI} and \ref{eq:transformsII}, we would expect all signal to be in the $\langle Q^{\prime} \rangle$ component if the polarization morphology were dominated by either a circular polarization morphology, expected for toroidal or radial magnetic fields, or a radial polarization morphology, expected for a poloidal magnetic field \citep[e.g.,][]{Lankhaar_ea_2021}. As we see no clear signal in the $Q^{\prime}$ component, but rather in the $\langle U^{\prime} \rangle$ component, we surmise that the \emph{azimuthally averaged} polarization structure must lie somewhere between the two. Figure~\ref{fig:azimuthal_dependence} shows how the values of $\langle Q^{\prime} \rangle$ and $\langle U^{\prime} \rangle$ vary depending on the underlying azimuthally symmetric polarization morphology. Assuming that $\langle Q^{\prime} \rangle = 0$, then $\langle U^{\prime} \rangle = \pm 1$, resulting in a spiral-like morphology shown in panels B and D in Fig.~\ref{fig:azimuthal_dependence}.

Under the assumption that only two polarization morphologies are present, clockwise spirals (i.e., spirals that are wound in a clockwise direction on the plane of the sky, $\langle U^{\prime} \rangle = +1$) and counter-clockwise spirals ($\langle U^{\prime} \rangle = -1$), it is possible to infer the $Q$ and $U$ morphologies consistent with the teardrop plots using Equations~\ref{eq:transformsI} and \ref{eq:transformsII}, and to then calculate $\chi$ using Eqn.~\ref{eq:polarization_angle}. Adopting a simple 2D model of the molecular line emission (i.e., a model in which every location in the disk can be described by a Gaussian emission line with a center that is shifted relative to the systemic velocity by the projected Keplerian rotation at that location), we can reconstruct a polarization morphology that is consistent with the teardrops, as shown in Fig.~\ref{fig:model_polarization}. In this figure, all polarization line segments are plotted with the same length to highlight the polarization morphology rather than any spatial variations in the polarization fraction.

A particularly intriguing aspect of the inferred polarization morphology is that $\langle U^{\prime} \rangle$ flips sign about the line center, such that the blue-shifted wing of the emission has a morphology described in panel B, while the red-shifted wing of the emission has the morphology shown in panel D. This results in a subtle difference in the polarization morphology between the North Eastern side of the disk and the South Western side. In the NE half of the disk, polarization orientations appear to align with the maximum gradient of $I$, while in the SW side of the disk, the orientations are perpendicular to the maximum gradient of $I$. Following the spiral morphology described in \citet{Teague_ea_2019a}, we assume that the former side is tilted \emph{away} from the observer, while the latter is tilted \emph{towards} the observer. Such a breaking in symmetry could be explained by the perceived difference in inclination of the disk on either side of the major axis. Given the low inclination of TW~Hya, the elevated emission surface of CO isotopologue emission \citep[$z / r \sim 0.3$, e.g.,][]{Pinte_ea_2018a} will make a considerable difference to the projected inclination of the disk. That is, along the minor axis the projected inclination is $i_{\rm proj} = \tan^{-1}(z / r) \pm i$, where $i_{\rm proj}$ is smaller for the NE side (tilted away from the observer), and larger $i_{\rm proj}$ for the SW side (tilted towards the observer). This difference in projected inclination could be the cause of the flip in polarization morphology, as the polarization morphology of the Goldreich-Kylafis effect is known to be sensitive to the geometrical properties \citep[e.g.,][]{Crutcher_2012}.

\citet{Lankhaar_Vlemmings_2020} and \citet{Lankhaar_ea_2021} presented synthetic observations for a toy model of a protoplanetary disk viewed at various inclinations. While the physical and chemical structure is not modeled after TW~Hya, the results provide context for the non-detections we report here. \citeauthor{Lankhaar_Vlemmings_2020} and \citeauthor{Lankhaar_ea_2021} considered three magnetic field morphologies: a toroidal field, a poloidal field, and a radial field and several different viewing geometries. For both the toroidal and radial magnetic field structures, the authors found an azimuthal polarization morphology, with polarization fractions $\sim 1\%$ in the line wings. When viewed face-on, the poloidal magnetic field morphology yielded a radial polarization morphology; however, at higher inclinations this transformed to a more circular morphology. Again, the polarization fractions were found to be $\sim 1\%$ in the line wings. These polarization fractions are consistent with those inferred for TW~Hya, although the data suggest a less ordered polarization morphology than was found for the toy model.

Future modeling efforts using a source-specific physical and chemical structure will be instrumental in guiding the analysis of the Stokes $Q$ and $U$ components. In particular, a strong prior on the $Q$ and $U$ morphologies will aid in the use of the shift-and-stack techniques employed here to tease out low-level signals.

\subsection{Circular Polarization}
\label{sec:discussion:circular_polarization}

\begin{figure}
    \centering
    \includegraphics[width=\columnwidth]{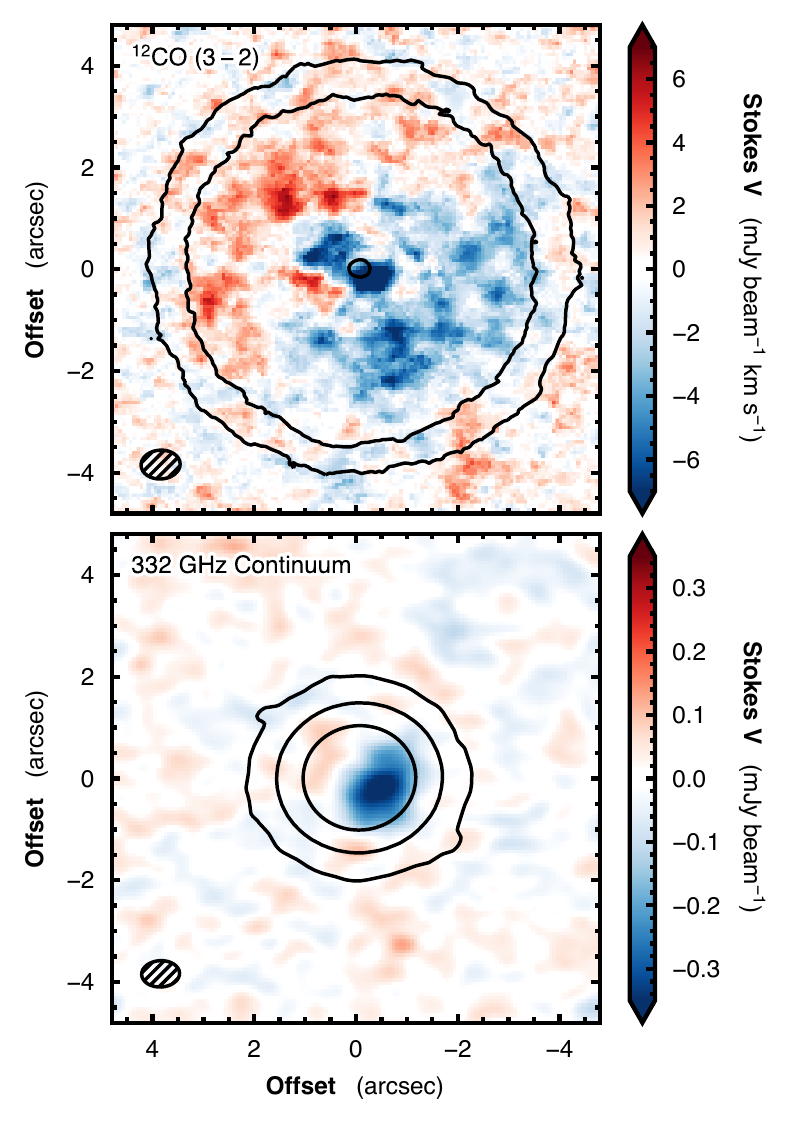}
    \caption{Comparison of the Stokes~$V$ components detected in the $^{12}$CO, \emph{top}, and the 332~GHz continuum, \emph{bottom}. The $^{12}$CO emission is integrated along the velocity axis, including the Stokes~$V$ emission where the absolute Stokes~$I$ component is greater than twice the background, $\sigma_{\rm I} = 4~{\rm mJy~beam^{-1}}$. In both panels, the three solid contours show signal-to-noise ratios in the Stokes~$I$ (integrated along the velocity axis in the case of the $^{12}$CO emission) component of 10, 100 and 1000. A clear dipole morphology is seen in both panels, highly suggestive that this signal is due to beam squint. The synthesized beam is shown in the bottom left of each panel.}
    \label{fig:stokes_v}
\end{figure}

For completeness, we report the detection of Stokes $V$ emission (circular polarization) in both the continuum and $^{12}$CO emission as shown in Fig.~\ref{fig:stokes_v}. Both detections share a similar morphology: a dipole pattern aligned along the minor axis of the disk such that the south-west half of the disk displays negative values, and the north-east side positive. In both cases the polarization fraction is $\sim 0.5\%$.

\citet{Houde_ea_2013} demonstrated that resonant scattering can transform linear polarization into circular polarization, providing a potential mechanisms for linear polarization to ``leak'' from the linear states into the circular. However, given the strong correlation in both polarization morphology and fraction between the $^{12}$CO and continuum emission, we attribute this signal to the well known instrumental effect of beam squint \citep[see, e.g.,][]{Hull2020b}. 

\section{Summary}
\label{sec:summary}

We present ALMA 332~GHz observations of polarized emission from the disk of TW~Hya, covering the $^{12}$CO (3-2), $^{13}$CO (3-2), and CS (7-6) transitions in addition to a continuum window at 332~GHz.

We detect linear polarization in the dust continuum emission, with polarization orientations that are approximately azimuthal. The median polarization fraction $P_{\rm frac}$ is 0.19\% in regions where the SNR of the polarized intensity $P$ is $>3$. We see a tentative increase in $P_{\rm frac}$ as a function of radius; however, we attribute this to the limited spatial resolution of the data. 

We compare our continuum polarization observations with lower frequency Band 6 observations, which also show an azimuthal morphology and have a similar polarization fraction. There may be a subtle difference between the azimuthal morphologies observed at the two different wavelengths, which could be indicative of changing polarization mechanisms between the two sets of observations; however, higher spatial resolution data are necessary to differentiate robustly the emission morphologies.

We do not detect significant linear polarization in the channel maps for any of the three molecular lines. However, by using an aligning and averaging approach used to detect weak line emission, and by adopting a linear combination of the $Q$ and $U$ components to account for potential azimuthal structure in their morphology, we detect linear polarization in the line wings of $^{12}$CO and $^{13}$CO.  The polarized emission arises between radii of $0\farcs5$ and $1\arcsec$ and reaches polarization fractions of $\sim 5\%$ and $\sim 3\%$ in $^{12}$CO and $^{13}$CO, respectively. We find these signals are significant at the $\sim 10\sigma$ and $\sim 5\sigma$ level for $^{12}$CO and $^{13}$CO, respectively for each wing. A similar polarization morphology is tentatively detected in the CS emission, albeit at a much less significant level ($\lesssim 3\sigma$).

We find that the sign of the polarization signal differs from the blue-shifted wing of the line to the red-shifted wing, suggesting a change in polarization morphology. Adopting a simple 2D analytical model for the disk emission morphology, and assuming two azimuthally symmetric polarization morphologies (one in each line wing), we are able to reconstruct a polarization morphology that was consistent with the detected signals. An asymmetry in the morphology across the major axis of the disk may arise due to the change in the projected inclination of the disk from the flared emission surface traced by $^{12}$CO emission.


\acknowledgments
The authors thank the anonymous referees for constructive reports.
This paper makes use of the following ALMA data: 2018.1.00980.S and 2018.1.00167.S. ALMA is a partnership of ESO (representing its member states), NSF (USA) and NINS (Japan), together with NRC (Canada), NSC and ASIAA (Taiwan), and KASI (Republic of Korea), in cooperation with the Republic of Chile. The Joint ALMA Observatory is operated by ESO, AUI/NRAO and NAOJ. The National Radio Astronomy Observatory is a facility of the National Science Foundation operated under cooperative agreement by Associated Universities, Inc.
R.T. acknowledges support from the Smithsonian Institution as a Submillimeter Array (SMA) Fellow. 
C.L.H.H. acknowledges the support of both the NAOJ Fellowship as well as JSPS KAKENHI grants 18K13586 and 20K14527.
T.H. acknowledges support from the European Research Council under the Horizon 2020 Framework Program via the ERC Advanced Grant Origins 83 24 28.
DS acknowledges support by the Deutsche Forschungsgemeinschaft through SPP 1833: ``Building a Habitable Earth'' (SE 1962/6-1).
RK acknowledges financial support via the Emmy Noether and Heisenberg Research Grants funded by the German Research Foundation (DFG) under grant no. KU 2849/3 and 2849/9.
This work was partly supported by the Programme National “Physique et Chimie du Milieu Interstellaire” (PCMI) of CNRS/INSU with INC/INP co-funded by CEA and CNES."
\bigskip \smallskip

\facilities{ALMA}
\software{\texttt{CASA} \citep{McMullin_ea_2007},
    \texttt{astropy} \citep{astropy_2013, astropy_2018},
    \texttt{GoFish} \citep{GoFish},
    \texttt{matplotlib} \citep{Hunter_2007},
    \texttt{scipy} \citep{scipy}.
}

\clearpage
\bibliography{bibliography}

\begin{thebibliography}{}
\expandafter\ifx\csname natexlab\endcsname\relax\def\natexlab#1{#1}\fi
\providecommand{\url}[1]{\href{#1}{#1}}
\providecommand{\dodoi}[1]{doi:~\href{http://doi.org/#1}{\nolinkurl{#1}}}
\providecommand{\doeprint}[1]{\href{http://ascl.net/#1}{\nolinkurl{http://ascl.net/#1}}}
\providecommand{\doarXiv}[1]{\href{https://arxiv.org/abs/#1}{\nolinkurl{https://arxiv.org/abs/#1}}}

\bibitem[{{Andersson} {et~al.}(2015){Andersson}, {Lazarian}, \&
  {Vaillancourt}}]{Andersson2015}
{Andersson}, B.-G., {Lazarian}, A., \& {Vaillancourt}, J.~E. 2015, \araa, 53,
  501, \dodoi{10.1146/annurev-astro-082214-122414}

\bibitem[{{Andrews} {et~al.}(2016){Andrews}, {Wilner}, {Zhu}, {Birnstiel},
  {Carpenter}, {P{\'e}rez}, {Bai}, {{\"O}berg}, {Hughes}, {Isella}, \&
  {Ricci}}]{Andrews_ea_2016}
{Andrews}, S.~M., {Wilner}, D.~J., {Zhu}, Z., {et~al.} 2016, \apjl, 820, L40,
  \dodoi{10.3847/2041-8205/820/2/L40}

\bibitem[{{Astropy Collaboration} {et~al.}(2013){Astropy Collaboration},
  {Robitaille}, {Tollerud}, {Greenfield}, {Droettboom}, {Bray}, {Aldcroft},
  {Davis}, {Ginsburg}, {Price-Whelan}, {Kerzendorf}, {Conley}, {Crighton},
  {Barbary}, {Muna}, {Ferguson}, {Grollier}, {Parikh}, {Nair}, {Unther},
  {Deil}, {Woillez}, {Conseil}, {Kramer}, {Turner}, {Singer}, {Fox}, {Weaver},
  {Zabalza}, {Edwards}, {Azalee Bostroem}, {Burke}, {Casey}, {Crawford},
  {Dencheva}, {Ely}, {Jenness}, {Labrie}, {Lim}, {Pierfederici}, {Pontzen},
  {Ptak}, {Refsdal}, {Servillat}, \& {Streicher}}]{astropy_2013}
{Astropy Collaboration}, {Robitaille}, T.~P., {Tollerud}, E.~J., {et~al.} 2013,
  \aap, 558, A33, \dodoi{10.1051/0004-6361/201322068}

\bibitem[{{Bacciotti} {et~al.}(2018){Bacciotti}, {Girart}, {Padovani}, {Podio},
  {Paladino}, {Testi}, {Bianchi}, {Galli}, {Codella}, {Coffey}, {Favre}, \&
  {Fedele}}]{Bacciotti2018}
{Bacciotti}, F., {Girart}, J.~M., {Padovani}, M., {et~al.} 2018, \apjl, 865,
  L12, \dodoi{10.3847/2041-8213/aadf87}

\bibitem[{{Bailer-Jones} {et~al.}(2018){Bailer-Jones}, {Rybizki}, {Fouesneau},
  {Mantelet}, \& {Andrae}}]{Bailer-Jones_ea_2018}
{Bailer-Jones}, C.~A.~L., {Rybizki}, J., {Fouesneau}, M., {Mantelet}, G., \&
  {Andrae}, R. 2018, \aj, 156, 58, \dodoi{10.3847/1538-3881/aacb21}

\bibitem[{{Balbus} \& {Hawley}(1991)}]{Balbus_Hawley_1991}
{Balbus}, S.~A., \& {Hawley}, J.~F. 1991, \apj, 376, 214,
  \dodoi{10.1086/170270}

\bibitem[{{Beuther} {et~al.}(2010){Beuther}, {Vlemmings}, {Rao}, \& {van der
  Tak}}]{Beuther_ea_2010}
{Beuther}, H., {Vlemmings}, W.~H.~T., {Rao}, R., \& {van der Tak}, F.~F.~S.
  2010, \apjl, 724, L113, \dodoi{10.1088/2041-8205/724/1/L113}

\bibitem[{{Blandford} \& {Payne}(1982)}]{Blandford_Payne_1982}
{Blandford}, R.~D., \& {Payne}, D.~G. 1982, \mnras, 199, 883,
  \dodoi{10.1093/mnras/199.4.883}

\bibitem[{{Boehler} {et~al.}(2017){Boehler}, {Weaver}, {Isella}, {Ricci},
  {Grady}, {Carpenter}, \& {Perez}}]{Boehler_ea_2017}
{Boehler}, Y., {Weaver}, E., {Isella}, A., {et~al.} 2017, \apj, 840, 60,
  \dodoi{10.3847/1538-4357/aa696c}

\bibitem[{{Ching} {et~al.}(2016){Ching}, {Lai}, {Zhang}, {Yang}, {Girart}, \&
  {Rao}}]{Ching_ea_2016}
{Ching}, T.-C., {Lai}, S.-P., {Zhang}, Q., {et~al.} 2016, \apj, 819, 159,
  \dodoi{10.3847/0004-637X/819/2/159}

\bibitem[{{Crutcher}(2012)}]{Crutcher_2012}
{Crutcher}, R.~M. 2012, \araa, 50, 29,
  \dodoi{10.1146/annurev-astro-081811-125514}

\bibitem[{{Dent} {et~al.}(2019){Dent}, {Pinte}, {Cortes}, {M{\'e}nard},
  {Hales}, {Fomalont}, \& {de Gregorio-Monsalvo}}]{Dent_ea_2019}
{Dent}, W.~R.~F., {Pinte}, C., {Cortes}, P.~C., {et~al.} 2019, \mnras, 482,
  L29, \dodoi{10.1093/mnrasl/sly181}

\bibitem[{{Flaherty} {et~al.}(2020){Flaherty}, {Hughes}, {Simon}, {Qi}, {Bai},
  {Bulatek}, {Andrews}, {Wilner}, \& {K{\'o}sp{\'a}l}}]{Flaherty_ea_2020}
{Flaherty}, K., {Hughes}, A.~M., {Simon}, J.~B., {et~al.} 2020, \apj, 895, 109,
  \dodoi{10.3847/1538-4357/ab8cc5}

\bibitem[{{Flaherty} {et~al.}(2015){Flaherty}, {Hughes}, {Rosenfeld},
  {Andrews}, {Chiang}, {Simon}, {Kerzner}, \& {Wilner}}]{Flaherty_ea_2015}
{Flaherty}, K.~M., {Hughes}, A.~M., {Rosenfeld}, K.~A., {et~al.} 2015, \apj,
  813, 99, \dodoi{10.1088/0004-637X/813/2/99}

\bibitem[{{Flaherty} {et~al.}(2017){Flaherty}, {Hughes}, {Rose}, {Simon}, {Qi},
  {Andrews}, {K{\'o}sp{\'a}l}, {Wilner}, {Chiang}, {Armitage}, \&
  {Bai}}]{Flaherty_ea_2017}
{Flaherty}, K.~M., {Hughes}, A.~M., {Rose}, S.~C., {et~al.} 2017, \apj, 843,
  150, \dodoi{10.3847/1538-4357/aa79f9}

\bibitem[{{Flock} {et~al.}(2017){Flock}, {Nelson}, {Turner}, {Bertrang},
  {Carrasco-Gonz{\'a}lez}, {Henning}, {Lyra}, \& {Teague}}]{Flock_ea_2017}
{Flock}, M., {Nelson}, R.~P., {Turner}, N.~J., {et~al.} 2017, \apj, 850, 131,
  \dodoi{10.3847/1538-4357/aa943f}

\bibitem[{{Flock} {et~al.}(2015){Flock}, {Ruge}, {Dzyurkevich}, {Henning},
  {Klahr}, \& {Wolf}}]{Flock_ea_2015}
{Flock}, M., {Ruge}, J.~P., {Dzyurkevich}, N., {et~al.} 2015, \aap, 574, A68,
  \dodoi{10.1051/0004-6361/201424693}

\bibitem[{{Gammie}(1996)}]{Gammie_1996}
{Gammie}, C.~F. 1996, \apj, 457, 355, \dodoi{10.1086/176735}

\bibitem[{{Girart} {et~al.}(1999){Girart}, {Crutcher}, \& {Rao}}]{Girart1999}
{Girart}, J.~M., {Crutcher}, R.~M., \& {Rao}, R. 1999, \apjl, 525, L109,
  \dodoi{10.1086/312345}

\bibitem[{{Gold}(1952)}]{Gold_1952}
{Gold}, T. 1952, \mnras, 112, 215, \dodoi{10.1093/mnras/112.2.215}

\bibitem[{{Goldreich} \& {Kylafis}(1981)}]{Goldreich_Kylafis_1981}
{Goldreich}, P., \& {Kylafis}, N.~D. 1981, \apjl, 243, L75,
  \dodoi{10.1086/183446}

\bibitem[{{Goldreich} \& {Kylafis}(1982)}]{Goldreich_Kylafis_1982}
---. 1982, \apj, 253, 606, \dodoi{10.1086/159663}

\bibitem[{{Guillet} {et~al.}(2020){Guillet}, {Girart}, {Maury}, \&
  {Alves}}]{Guillet_ea_2020}
{Guillet}, V., {Girart}, J.~M., {Maury}, A.~J., \& {Alves}, F.~O. 2020, \aap,
  634, L15, \dodoi{10.1051/0004-6361/201937314}

\bibitem[{{Guilloteau} {et~al.}(2012){Guilloteau}, {Dutrey}, {Wakelam},
  {Hersant}, {Semenov}, {Chapillon}, {Henning}, \&
  {Pi{\'e}tu}}]{Guilloteau_ea_2012}
{Guilloteau}, S., {Dutrey}, A., {Wakelam}, V., {et~al.} 2012, \aap, 548, A70,
  \dodoi{10.1051/0004-6361/201220331}

\bibitem[{{Guzm{\'a}n} {et~al.}(2015){Guzm{\'a}n}, {{\"O}berg}, {Loomis}, \&
  {Qi}}]{Guzman_ea_2015}
{Guzm{\'a}n}, V.~V., {{\"O}berg}, K.~I., {Loomis}, R., \& {Qi}, C. 2015, \apj,
  814, 53, \dodoi{10.1088/0004-637X/814/1/53}

\bibitem[{{Harrison} {et~al.}(2019){Harrison}, {Looney}, {Stephens}, {Li},
  {Yang}, {Kataoka}, {Harris}, {Kwon}, {Muto}, \& {Momose}}]{Harrison_ea_2019}
{Harrison}, R.~E., {Looney}, L.~W., {Stephens}, I.~W., {et~al.} 2019, \apjl,
  877, L2, \dodoi{10.3847/2041-8213/ab1e46}

\bibitem[{{Harrison} {et~al.}(2021){Harrison}, {Looney}, {Stephens}, {Li},
  {Teague}, {Crutcher}, {Yang}, {Cox}, {Fern{\'a}ndez-L{\'o}pez}, \&
  {Shinnaga}}]{Harrison_ea_2021}
---. 2021, \apj, 908, 141, \dodoi{10.3847/1538-4357/abd94e}

\bibitem[{{Houde} {et~al.}(2013){Houde}, {Hezareh}, {Jones}, \&
  {Rajabi}}]{Houde_ea_2013}
{Houde}, M., {Hezareh}, T., {Jones}, S., \& {Rajabi}, F. 2013, \apj, 764, 24,
  \dodoi{10.1088/0004-637X/764/1/24}

\bibitem[{{Huang} {et~al.}(2018){Huang}, {Andrews}, {Cleeves}, {{\"O}berg},
  {Wilner}, {Bai}, {Birnstiel}, {Carpenter}, {Hughes}, {Isella}, {P{\'e}rez},
  {Ricci}, \& {Zhu}}]{Huang_ea_2018}
{Huang}, J., {Andrews}, S.~M., {Cleeves}, L.~I., {et~al.} 2018, \apj, 852, 122,
  \dodoi{10.3847/1538-4357/aaa1e7}

\bibitem[{{Hughes} {et~al.}(2011){Hughes}, {Wilner}, {Andrews}, {Qi}, \&
  {Hogerheijde}}]{Hughes_ea_2011}
{Hughes}, A.~M., {Wilner}, D.~J., {Andrews}, S.~M., {Qi}, C., \& {Hogerheijde},
  M.~R. 2011, \apj, 727, 85, \dodoi{10.1088/0004-637X/727/2/85}

\bibitem[{{Hull} \& {Plambeck}(2015)}]{Hull_Plambeck_2015}
{Hull}, C. L.~H., \& {Plambeck}, R.~L. 2015, Journal of Astronomical
  Instrumentation, 4, 1550005, \dodoi{10.1142/S2251171715500051}

\bibitem[{{Hull} \& {Zhang}(2019)}]{Hull_Zhang_2019}
{Hull}, C. L.~H., \& {Zhang}, Q. 2019, Frontiers in Astronomy and Space
  Sciences, 6, 3, \dodoi{10.3389/fspas.2019.00003}

\bibitem[{{Hull} {et~al.}(2018){Hull}, {Yang}, {Li}, {Kataoka}, {Stephens},
  {Andrews}, {Bai}, {Cleeves}, {Hughes}, {Looney}, {P{\'e}rez}, \&
  {Wilner}}]{Hull_ea_2018}
{Hull}, C. L.~H., {Yang}, H., {Li}, Z.-Y., {et~al.} 2018, \apj, 860, 82,
  \dodoi{10.3847/1538-4357/aabfeb}

\bibitem[{{Hull} {et~al.}(2020){Hull}, {Cortes}, {Gouellec}, {Girart}, {Nagai},
  {Nakanishi}, {Kameno}, {Fomalont}, {Brogan}, {Moellenbrock}, {Paladino}, \&
  {Villard}}]{Hull2020b}
{Hull}, C. L.~H., {Cortes}, P.~C., {Gouellec}, V. J.~M.~L., {et~al.} 2020,
  \pasp, 132, 094501, \dodoi{10.1088/1538-3873/ab99cd}

\bibitem[{Hunter(2007)}]{Hunter_2007}
Hunter, J.~D. 2007, Computing in Science \& Engineering, 9, 90,
  \dodoi{10.1109/MCSE.2007.55}

\bibitem[{{Kataoka} {et~al.}(2019){Kataoka}, {Okuzumi}, \&
  {Tazaki}}]{Kataoka_ea_2019}
{Kataoka}, A., {Okuzumi}, S., \& {Tazaki}, R. 2019, \apjl, 874, L6,
  \dodoi{10.3847/2041-8213/ab0c9a}

\bibitem[{{Kataoka} {et~al.}(2017){Kataoka}, {Tsukagoshi}, {Pohl}, {Muto},
  {Nagai}, {Stephens}, {Tomisaka}, \& {Momose}}]{Kataoka2017}
{Kataoka}, A., {Tsukagoshi}, T., {Pohl}, A., {et~al.} 2017, \apjl, 844, L5,
  \dodoi{10.3847/2041-8213/aa7e33}

\bibitem[{{Kataoka} {et~al.}(2015){Kataoka}, {Muto}, {Momose}, {Tsukagoshi},
  {Fukagawa}, {Shibai}, {Hanawa}, {Murakawa}, \& {Dullemond}}]{Kataoka_ea_2015}
{Kataoka}, A., {Muto}, T., {Momose}, M., {et~al.} 2015, \apj, 809, 78,
  \dodoi{10.1088/0004-637X/809/1/78}

\bibitem[{{Kataoka} {et~al.}(2016){Kataoka}, {Tsukagoshi}, {Momose}, {Nagai},
  {Muto}, {Dullemond}, {Pohl}, {Fukagawa}, {Shibai}, {Hanawa}, \&
  {Murakawa}}]{Kataoka_ea_2016}
{Kataoka}, A., {Tsukagoshi}, T., {Momose}, M., {et~al.} 2016, \apjl, 831, L12,
  \dodoi{10.3847/2041-8205/831/2/L12}

\bibitem[{{Killeen} {et~al.}(1986){Killeen}, {Bicknell}, \&
  {Ekers}}]{Killeen_ea_1986}
{Killeen}, N.~E.~B., {Bicknell}, G.~V., \& {Ekers}, R.~D. 1986, \apj, 302, 306,
  \dodoi{10.1086/163992}

\bibitem[{{Lankhaar} \& {Vlemmings}(2020)}]{Lankhaar_Vlemmings_2020}
{Lankhaar}, B., \& {Vlemmings}, W. 2020, \aap, 636, A14,
  \dodoi{10.1051/0004-6361/202037509}

\bibitem[{{Lankhaar} {et~al.}(2021){Lankhaar}, {Vlemmings}, \&
  {Bjerkeli}}]{Lankhaar_ea_2021}
{Lankhaar}, B., {Vlemmings}, W., \& {Bjerkeli}, P. 2021, arXiv e-prints,
  arXiv:2105.06482.
\newblock \doarXiv{2105.06482}

\bibitem[{{Lazarian} \& {Hoang}(2007)}]{LazarianHoang2007a}
{Lazarian}, A., \& {Hoang}, T. 2007, \mnras, 378, 910,
  \dodoi{10.1111/j.1365-2966.2007.11817.x}

\bibitem[{{Lee} {et~al.}(2018){Lee}, {Hwang}, {Ching}, {Hirano}, {Lai}, {Rao},
  \& {Ho}}]{Lee_ea_2018}
{Lee}, C.-F., {Hwang}, H.-C., {Ching}, T.-C., {et~al.} 2018, Nature
  Communications, 9, 4636, \dodoi{10.1038/s41467-018-07143-8}

\bibitem[{{Mac{\'\i}as} {et~al.}(2021){Mac{\'\i}as}, {Guerra-Alvarado},
  {Carrasco-Gonz{\'a}lez}, {Ribas}, {Espaillat}, {Huang}, \&
  {Andrews}}]{Macias_ea_2021}
{Mac{\'\i}as}, E., {Guerra-Alvarado}, O., {Carrasco-Gonz{\'a}lez}, C., {et~al.}
  2021, \aap, 648, A33, \dodoi{10.1051/0004-6361/202039812}

\bibitem[{{Matr{\`a}} {et~al.}(2017){Matr{\`a}}, {MacGregor}, {Kalas}, {Wyatt},
  {Kennedy}, {Wilner}, {Duchene}, {Hughes}, {Pan}, {Shannon}, {Clampin},
  {Fitzgerald}, {Graham}, {Holland }, {Pani{\'c}}, \& {Su}}]{Matra_ea_2017}
{Matr{\`a}}, L., {MacGregor}, M.~A., {Kalas}, P., {et~al.} 2017, \apj, 842, 9,
  \dodoi{10.3847/1538-4357/aa71b4}

\bibitem[{{McMullin} {et~al.}(2007){McMullin}, {Waters}, {Schiebel}, {Young},
  \& {Golap}}]{McMullin_ea_2007}
{McMullin}, J.~P., {Waters}, B., {Schiebel}, D., {Young}, W., \& {Golap}, K.
  2007, Astronomical Society of the Pacific Conference Series, Vol. 376, {CASA
  Architecture and Applications}, ed. R.~A. {Shaw}, F.~{Hill}, \& D.~J. {Bell},
  127

\bibitem[{{Mori} {et~al.}(2019){Mori}, {Kataoka}, {Ohashi}, {Momose}, {Muto},
  {Nagai}, \& {Tsukagoshi}}]{Mori2019}
{Mori}, T., {Kataoka}, A., {Ohashi}, S., {et~al.} 2019, \apj, 883, 16,
  \dodoi{10.3847/1538-4357/ab3575}

\bibitem[{{Morris} {et~al.}(1985){Morris}, {Lucas}, \&
  {Omont}}]{Morris_ea_1985}
{Morris}, M., {Lucas}, R., \& {Omont}, A. 1985, \aap, 142, 107

\bibitem[{{Nagai} {et~al.}(2016){Nagai}, {Nakanishi}, {Paladino}, {Hull},
  {Cortes}, {Moellenbrock}, {Fomalont}, {Asada}, \& {Hada}}]{Nagai2016}
{Nagai}, H., {Nakanishi}, K., {Paladino}, R., {et~al.} 2016, \apj, 824, 132,
  \dodoi{10.3847/0004-637X/824/2/132}

\bibitem[{{Ohashi} {et~al.}(2018){Ohashi}, {Kataoka}, {Nagai}, {Momose},
  {Muto}, {Hanawa}, {Fukagawa}, {Tsukagoshi}, {Murakawa}, \&
  {Shibai}}]{Ohashi_ea_2018}
{Ohashi}, S., {Kataoka}, A., {Nagai}, H., {et~al.} 2018, \apj, 864, 81,
  \dodoi{10.3847/1538-4357/aad632}

\bibitem[{{Pattle} \& {Fissel}(2019)}]{PattleFissel2019}
{Pattle}, K., \& {Fissel}, L. 2019, Frontiers in Astronomy and Space Sciences,
  6, 15, \dodoi{10.3389/fspas.2019.00015}

\bibitem[{{Pinte} {et~al.}(2018){Pinte}, {M{\'e}nard}, {Duch{\^e}ne}, {Hill},
  {Dent}, {Woitke}, {Maret}, {van der Plas}, {Hales}, {Kamp}, {Thi}, {de
  Gregorio-Monsalvo}, {Rab}, {Quanz}, {Avenhaus}, {Carmona}, \&
  {Casassus}}]{Pinte_ea_2018a}
{Pinte}, C., {M{\'e}nard}, F., {Duch{\^e}ne}, G., {et~al.} 2018, \aap, 609,
  A47, \dodoi{10.1051/0004-6361/201731377}

\bibitem[{{Planck Collaboration} {et~al.}(2016){Planck Collaboration}, {Ade},
  {Aghanim}, {Alves}, {Arnaud}, {Arzoumanian}, {Ashdown}, {Aumont},
  {Baccigalupi}, {Banday}, {Barreiro}, {Bartolo}, {Battaner}, {Benabed},
  {Beno{\^\i}t}, {Benoit-L{\'e}vy}, {Bernard}, {Bersanelli}, {Bielewicz},
  {Bock}, {Bonavera}, {Bond}, {Borrill}, {Bouchet}, {Boulanger}, {Bracco},
  {Burigana}, {Calabrese}, {Cardoso}, {Catalano}, {Chiang}, {Christensen},
  {Colombo}, {Combet}, {Couchot}, {Crill}, {Curto}, {Cuttaia}, {Danese},
  {Davies}, {Davis}, {de Bernardis}, {de Rosa}, {de Zotti}, {Delabrouille},
  {Dickinson}, {Diego}, {Dole}, {Donzelli}, {Dor{\'e}}, {Douspis}, {Ducout},
  {Dupac}, {Efstathiou}, {Elsner}, {En{\ss}lin}, {Eriksen}, {Falceta-Gon{\c
  c}alves}, {Falgarone}, {Ferri{\`e}re}, {Finelli}, {Forni}, {Frailis},
  {Fraisse}, {Franceschi}, {Frejsel}, {Galeotta}, {Galli}, {Ganga}, {Ghosh},
  {Giard}, {Gjerl{\o}w}, {Gonz{\'a}lez-Nuevo}, {G{\'o}rski}, {Gregorio},
  {Gruppuso}, {Gudmundsson}, {Guillet}, {Harrison}, {Helou}, {Hennebelle},
  {Henrot-Versill{\'e}}, {Hern{\'a}ndez-Monteagudo}, {Herranz}, {Hildebrandt},
  {Hivon}, {Holmes}, {Hornstrup}, {Huffenberger}, {Hurier}, {Jaffe}, {Jaffe},
  {Jones}, {Juvela}, {Keih{\"a}nen}, {Keskitalo}, {Kisner}, {Knoche}, {Kunz},
  {Kurki-Suonio}, {Lagache}, {Lamarre}, {Lasenby}, {Lattanzi}, {Lawrence},
  {Leonardi}, {Levrier}, {Liguori}, {Lilje}, {Linden-V{\o}rnle},
  {L{\'o}pez-Caniego}, {Lubin}, {Mac{\'{\i}}as-P{\'e}rez}, {Maino},
  {Mandolesi}, {Mangilli}, {Maris}, {Martin}, {Mart{\'{\i}}nez-Gonz{\'a}lez},
  {Masi}, {Matarrese}, {Melchiorri}, {Mendes}, {Mennella}, {Migliaccio},
  {Miville-Desch{\^e}nes}, {Moneti}, {Montier}, {Morgante}, {Mortlock},
  {Munshi}, {Murphy}, {Naselsky}, {Nati}, {Netterfield}, {Noviello}, {Novikov},
  {Novikov}, {Oppermann}, {Oxborrow}, {Pagano}, {Pajot}, {Paladini},
  {Paoletti}, {Pasian}, {Perotto}, {Pettorino}, {Piacentini}, {Piat},
  {Pierpaoli}, {Pietrobon}, {Plaszczynski}, {Pointecouteau}, {Polenta},
  {Ponthieu}, {Pratt}, {Prunet}, {Puget}, {Rachen}, {Reinecke}, {Remazeilles},
  {Renault}, {Renzi}, {Ristorcelli}, {Rocha}, {Rossetti}, {Roudier},
  {Rubi{\~n}o-Mart{\'{\i}}n}, {Rusholme}, {Sandri}, {Santos}, {Savelainen},
  {Savini}, {Scott}, {Soler}, {Stolyarov}, {Sudiwala}, {Sutton}, {Suur-Uski},
  {Sygnet}, {Tauber}, {Terenzi}, {Toffolatti}, {Tomasi}, {Tristram}, {Tucci},
  {Umana}, {Valenziano}, {Valiviita}, {Van Tent}, {Vielva}, {Villa}, {Wade},
  {Wandelt}, {Wehus}, {Ysard}, {Yvon}, \& {Zonca}}]{PlanckXXXV}
{Planck Collaboration}, {Ade}, P.~A.~R., {Aghanim}, N., {et~al.} 2016, \aap,
  586, A138, \dodoi{10.1051/0004-6361/201525896}

\bibitem[{{Pohl} {et~al.}(2016){Pohl}, {Kataoka}, {Pinilla}, {Dullemond},
  {Henning}, \& {Birnstiel}}]{Pohl_ea_2016}
{Pohl}, A., {Kataoka}, A., {Pinilla}, P., {et~al.} 2016, \aap, 593, A12,
  \dodoi{10.1051/0004-6361/201628637}

\bibitem[{{Price-Whelan} {et~al.}(2018){Price-Whelan}, {Sip{\H{o}}cz},
  {G{\"u}nther}, {Lim}, {Crawford}, {Conseil}, {Shupe}, {Craig}, {Dencheva},
  {Ginsburg}, {VanderPlas}, {Bradley}, {P{\'e}rez-Su{\'a}rez}, {de Val-Borro},
  {Paper Contributors}, {Aldcroft}, {Cruz}, {Robitaille}, {Tollerud},
  {Coordination Committee}, {Ardelean}, {Babej}, {Bach}, {Bachetti}, {Bakanov},
  {Bamford}, {Barentsen}, {Barmby}, {Baumbach}, {Berry}, {Biscani}, {Boquien},
  {Bostroem}, {Bouma}, {Brammer}, {Bray}, {Breytenbach}, {Buddelmeijer},
  {Burke}, {Calderone}, {Cano Rodr{\'\i}guez}, {Cara}, {Cardoso}, {Cheedella},
  {Copin}, {Corrales}, {Crichton}, {D{\textquoteright}Avella}, {Deil},
  {Depagne}, {Dietrich}, {Donath}, {Droettboom}, {Earl}, {Erben}, {Fabbro},
  {Ferreira}, {Finethy}, {Fox}, {Garrison}, {Gibbons}, {Goldstein}, {Gommers},
  {Greco}, {Greenfield}, {Groener}, {Grollier}, {Hagen}, {Hirst}, {Homeier},
  {Horton}, {Hosseinzadeh}, {Hu}, {Hunkeler}, {Ivezi{\'c}}, {Jain}, {Jenness},
  {Kanarek}, {Kendrew}, {Kern}, {Kerzendorf}, {Khvalko}, {King}, {Kirkby},
  {Kulkarni}, {Kumar}, {Lee}, {Lenz}, {Littlefair}, {Ma}, {Macleod},
  {Mastropietro}, {McCully}, {Montagnac}, {Morris}, {Mueller}, {Mumford},
  {Muna}, {Murphy}, {Nelson}, {Nguyen}, {Ninan}, {N{\"o}the}, {Ogaz}, {Oh},
  {Parejko}, {Parley}, {Pascual}, {Patil}, {Patil}, {Plunkett}, {Prochaska},
  {Rastogi}, {Reddy Janga}, {Sabater}, {Sakurikar}, {Seifert}, {Sherbert},
  {Sherwood-Taylor}, {Shih}, {Sick}, {Silbiger}, {Singanamalla}, {Singer},
  {Sladen}, {Sooley}, {Sornarajah}, {Streicher}, {Teuben}, {Thomas},
  {Tremblay}, {Turner}, {Terr{\'o}n}, {van Kerkwijk}, {de la Vega}, {Watkins},
  {Weaver}, {Whitmore}, {Woillez}, {Zabalza}, \& {Contributors}}]{astropy_2018}
{Price-Whelan}, A.~M., {Sip{\H{o}}cz}, B.~M., {G{\"u}nther}, H.~M., {et~al.}
  2018, \aj, 156, 123, \dodoi{10.3847/1538-3881/aabc4f}

\bibitem[{{Remijan} {et~al.}(2020){Remijan}, {Biggs}, {Cortes}, {Dent}, {Di
  Francesco}, {Fomalont}, {Hales}, {Kameno}, {Mason}, {Philips}, {Saini},
  {Stoehr}, {Vila Vilaro}, \& {Villard}}]{ALMATechnicalHandbook}
{Remijan}, A., {Biggs}, A., {Cortes}, P., {et~al.} 2020, ALMA Technical
  Handbook, ALMA Doc. 8.3

\bibitem[{{Rosenfeld} {et~al.}(2012){Rosenfeld}, {Qi}, {Andrews}, {Wilner},
  {Corder}, {Dullemond}, {Lin}, {Hughes}, {D'Alessio}, \&
  {Ho}}]{Rosenfeld_ea_2012}
{Rosenfeld}, K.~A., {Qi}, C., {Andrews}, S.~M., {et~al.} 2012, \apj, 757, 129,
  \dodoi{10.1088/0004-637X/757/2/129}

\bibitem[{{Schwarz} {et~al.}(2016){Schwarz}, {Bergin}, {Cleeves}, {Blake},
  {Zhang}, {{\"O}berg}, {van Dishoeck}, \& {Qi}}]{Schwarz_ea_2016}
{Schwarz}, K.~R., {Bergin}, E.~A., {Cleeves}, L.~I., {et~al.} 2016, \apj, 823,
  91, \dodoi{10.3847/0004-637X/823/2/91}

\bibitem[{{Schwarz} {et~al.}(2019){Schwarz}, {Teague}, \&
  {Bergin}}]{Schwarz_ea_2019}
{Schwarz}, K.~R., {Teague}, R., \& {Bergin}, E.~A. 2019, \apjl, 876, L13,
  \dodoi{10.3847/2041-8213/ab1b0d}

\bibitem[{{Simmons} \& {Stewart}(1985)}]{Simmons_Stewart_1985}
{Simmons}, J.~F.~L., \& {Stewart}, B.~G. 1985, \aap, 142, 100

\bibitem[{{Stephens} {et~al.}(2020){Stephens}, {Fern{\'a}ndez-L{\'o}pez}, {Li},
  {Looney}, \& {Teague}}]{Stephens_ea_2020}
{Stephens}, I.~W., {Fern{\'a}ndez-L{\'o}pez}, M., {Li}, Z.-Y., {Looney}, L.~W.,
  \& {Teague}, R. 2020, \apj, 901, 71, \dodoi{10.3847/1538-4357/abaef7}

\bibitem[{{Stephens} {et~al.}(2014){Stephens}, {Looney}, {Kwon},
  {Fern{\'a}ndez-L{\'o}pez}, {Hughes}, {Mundy}, {Crutcher}, {Li}, \&
  {Rao}}]{Stephens_ea_2014}
{Stephens}, I.~W., {Looney}, L.~W., {Kwon}, W., {et~al.} 2014, \nat, 514, 597,
  \dodoi{10.1038/nature13850}

\bibitem[{{Stephens} {et~al.}(2017){Stephens}, {Yang}, {Li}, {Looney},
  {Kataoka}, {Kwon}, {Fern{\'a}ndez-L{\'o}pez}, {Hull}, {Hughes}, {Segura-Cox},
  {Mundy}, {Crutcher}, \& {Rao}}]{Stephens_ea_2017}
{Stephens}, I.~W., {Yang}, H., {Li}, Z.-Y., {et~al.} 2017, \apj, 851, 55,
  \dodoi{10.3847/1538-4357/aa998b}

\bibitem[{{Suriano} {et~al.}(2018){Suriano}, {Li}, {Krasnopolsky}, \&
  {Shang}}]{Suriano_ea_2018}
{Suriano}, S.~S., {Li}, Z.-Y., {Krasnopolsky}, R., \& {Shang}, H. 2018, \mnras,
  477, 1239, \dodoi{10.1093/mnras/sty717}

\bibitem[{{Tazaki} {et~al.}(2017){Tazaki}, {Lazarian}, \&
  {Nomura}}]{Tazaki_ea_2017}
{Tazaki}, R., {Lazarian}, A., \& {Nomura}, H. 2017, \apj, 839, 56,
  \dodoi{10.3847/1538-4357/839/1/56}

\bibitem[{Teague(2019)}]{GoFish}
Teague, R. 2019, The Journal of Open Source Software, 4, 1632,
  \dodoi{10.21105/joss.01632}

\bibitem[{{Teague} {et~al.}(2018{\natexlab{a}}){Teague}, {Bae}, {Bergin},
  {Birnstiel}, \& {Foreman-Mackey}}]{Teague_ea_2018a}
{Teague}, R., {Bae}, J., {Bergin}, E.~A., {Birnstiel}, T., \& {Foreman-Mackey},
  D. 2018{\natexlab{a}}, \apjl, 860, L12, \dodoi{10.3847/2041-8213/aac6d7}

\bibitem[{{Teague} {et~al.}(2019){Teague}, {Bae}, {Huang}, \&
  {Bergin}}]{Teague_ea_2019a}
{Teague}, R., {Bae}, J., {Huang}, J., \& {Bergin}, E.~A. 2019, \apjl, 884, L56,
  \dodoi{10.3847/2041-8213/ab4a83}

\bibitem[{{Teague} {et~al.}(2016){Teague}, {Guilloteau}, {Semenov}, {Henning},
  {Dutrey}, {Pi{\'e}tu}, {Birnstiel}, {Chapillon}, {Hollenbach}, \&
  {Gorti}}]{Teague_ea_2016}
{Teague}, R., {Guilloteau}, S., {Semenov}, D., {et~al.} 2016, \aap, 592, A49,
  \dodoi{10.1051/0004-6361/201628550}

\bibitem[{{Teague} {et~al.}(2018{\natexlab{b}}){Teague}, {Henning},
  {Guilloteau}, {Bergin}, {Semenov}, {Dutrey}, {Flock}, {Gorti}, \&
  {Birnstiel}}]{Teague_ea_2018b}
{Teague}, R., {Henning}, T., {Guilloteau}, S., {et~al.} 2018{\natexlab{b}},
  \apj, 864, 133, \dodoi{10.3847/1538-4357/aad80e}

\bibitem[{{Turner} {et~al.}(2014){Turner}, {Fromang}, {Gammie}, {Klahr},
  {Lesur}, {Wardle}, \& {Bai}}]{Turner_ea_2014}
{Turner}, N.~J., {Fromang}, S., {Gammie}, C., {et~al.} 2014, in Protostars and
  Planets VI, ed. H.~{Beuther}, R.~S. {Klessen}, C.~P. {Dullemond}, \&
  T.~{Henning}, 411, \dodoi{10.2458/azu_uapress_9780816531240-ch018}

\bibitem[{{Vaillancourt}(2006)}]{Vaillancourt_2006}
{Vaillancourt}, J.~E. 2006, \pasp, 118, 1340, \dodoi{10.1086/507472}

\bibitem[{{Virtanen} {et~al.}(2020){Virtanen}, {Gommers}, {Oliphant},
  {Haberland}, {Reddy}, {Cournapeau}, {Burovski}, {Peterson}, {Weckesser},
  {Bright}, {van der Walt}, {Brett}, {Wilson}, {Jarrod Millman}, {Mayorov},
  {Nelson}, {Jones}, {Kern}, {Larson}, {Carey}, {Polat}, {Feng}, {Moore}, {Vand
  erPlas}, {Laxalde}, {Perktold}, {Cimrman}, {Henriksen}, {Quintero}, {Harris},
  {Archibald}, {Ribeiro}, {Pedregosa}, {van Mulbregt}, \&
  {Contributors}}]{scipy}
{Virtanen}, P., {Gommers}, R., {Oliphant}, T.~E., {et~al.} 2020, Nature
  Methods, \dodoi{https://doi.org/10.1038/s41592-019-0686-2}

\bibitem[{{Vlemmings} {et~al.}(2017){Vlemmings}, {Khouri}, {Mart{\'\i}-Vidal},
  {Tafoya}, {Baudry}, {Etoka}, {Humphreys}, {Jones}, {Kemball}, {O'Gorman},
  {P{\'e}rez-S{\'a}nchez}, \& {Richards}}]{Vlemmings_ea_2017}
{Vlemmings}, W.~H.~T., {Khouri}, T., {Mart{\'\i}-Vidal}, I., {et~al.} 2017,
  \aap, 603, A92, \dodoi{10.1051/0004-6361/201730735}

\bibitem[{{Vlemmings} {et~al.}(2019){Vlemmings}, {Lankhaar}, {Cazzoletti},
  {Ceccobello}, {Dall'Olio}, {van Dishoeck}, {Facchini}, {Humphreys},
  {Persson}, {Testi}, \& {Williams}}]{Vlemmings_ea_2019}
{Vlemmings}, W.~H.~T., {Lankhaar}, B., {Cazzoletti}, P., {et~al.} 2019, \aap,
  624, L7, \dodoi{10.1051/0004-6361/201935459}

\bibitem[{{Wardle} \& {Kronberg}(1974)}]{Wardle_Kronberg_1974}
{Wardle}, J.~F.~C., \& {Kronberg}, P.~P. 1974, \apj, 194, 249,
  \dodoi{10.1086/153240}

\bibitem[{{Wurster} \& {Li}(2018)}]{Wurster_Li_2018}
{Wurster}, J., \& {Li}, Z.-Y. 2018, Frontiers in Astronomy and Space Sciences,
  5, 39, \dodoi{10.3389/fspas.2018.00039}

\bibitem[{{Yang} {et~al.}(2016){Yang}, {Li}, {Looney}, \&
  {Stephens}}]{Yang_ea_2016}
{Yang}, H., {Li}, Z.-Y., {Looney}, L., \& {Stephens}, I. 2016, \mnras, 456,
  2794, \dodoi{10.1093/mnras/stv2633}

\bibitem[{{Yang} {et~al.}(2019){Yang}, {Li}, {Stephens}, {Kataoka}, \&
  {Looney}}]{Yang_ea_2019}
{Yang}, H., {Li}, Z.-Y., {Stephens}, I.~W., {Kataoka}, A., \& {Looney}, L.
  2019, \mnras, 483, 2371, \dodoi{10.1093/mnras/sty3263}

\bibitem[{{Yen} {et~al.}(2016){Yen}, {Koch}, {Liu}, {Puspitaningrum}, {Hirano},
  {Lee}, \& {Takakuwa}}]{Yen_ea_2016}
{Yen}, H.-W., {Koch}, P.~M., {Liu}, H.~B., {et~al.} 2016, \apj, 832, 204,
  \dodoi{10.3847/0004-637X/832/2/204}

\end{thebibliography}
\bibliographystyle{aasjournal}

\appendix

\section{Debiasing Polarization Values}
\label{sec:app:debiasing}

The observed polarization $P$ is given by the quadrature sum of the Stokes $Q$ and $U$ components: $P = \sqrt{Q^2 + U^2}$. As $P$ is always positive, any noise in $Q$ or $U$ will result in an over-estimate of the true polarization signal $P_0$. There is a wealth of literature discussing this issue, and many approaches have been developed to correct for this bias and to estimate uncertainties in the inferred $P_0$ value. In this Appendix, we briefly discuss the problem, and outline our approach to debiasing and estimating uncertainties.

We first make the assumption that both the $Q$ and $U$ components are described by normal distributions: $Q \in \mathcal{N}(Q_0,\, \sigma)$ and $U \in \mathcal{N}(U_0,\, \sigma)$, where both distributions are described by the same variance, $\sigma^2$. Based on the results described in Section~\ref{sec:searching:CDFs}, this is an appropriate assumption. It can then be shown that the distribution of $P$ follows a Rice distribution:

\begin{equation}
    P \sim F(P|P_0,\sigma) = \frac{P}{\sigma} I_0 \left( \frac{PP_0}{\sigma^2} \right) \exp \left( - \frac{P^2 + P_0^2}{2 \sigma^2} \right), 
    \label{eq:app:rice}
\end{equation}

\noindent where $I_0$ is the zeroth-order modified Bessel function \citep[e.g.,][]{Simmons_Stewart_1985, Vaillancourt_2006}. Thus, for a given observed $\{P,\,\sigma\}$, it is possible to construct a probability distribution function for $P_0$ from which the most likely value of $P_0$ can be inferred after making some assumptions.

\begin{figure}
    \centering
    \includegraphics{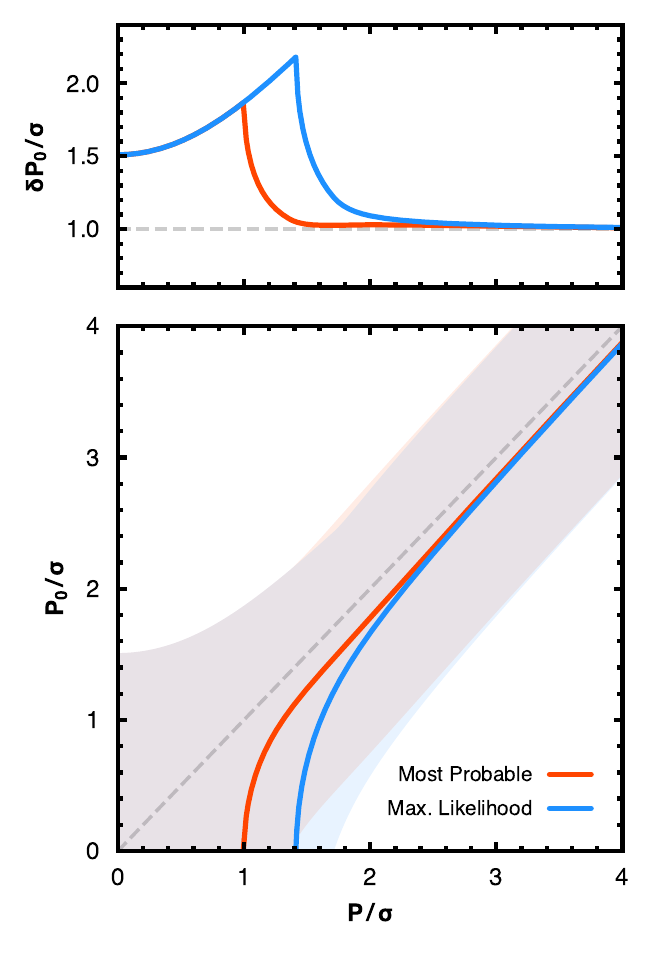}
    \caption{\emph{Top:} estimated uncertainty in $P_0$ compared with the uncertainty in $P$ from the observations. For large $P \, / \, \sigma$ values, this uncertainty approaches $\delta P_0 \approx \sigma$. The sharp increase at $P \, / \, \sigma \approx 1.4$ for the maximum likelihood estimator and $\approx 1.0$ for the most probable estimator arises because only upper limits can be calculated. \emph{Bottom:} the inferred $P_0$ value for a given $P$ value. For both estimators, the solid line shows the inferred $P_0$ that maximizes the likelihood, and the shaded region shows the $\pm \delta P_0$ region as defined in Equation \ref{eqn:uncertainty}.}
    \label{fig:app:debias_uncertainty}
\end{figure}

\citet{Simmons_Stewart_1985} explored several different estimators that can be used to infer $P_0$ from the probability distribution function; we refer the interested reader to this work for a more thorough discussion. Here we opt to use the ``most probable'' estimator described in \citet{Wardle_Kronberg_1974}, rather than the ``maximum likelihood'' estimator used in other works \citep[e.g.,][]{Vaillancourt_2006}, for reasons we describe later. The most probable estimator returns the $\hat{P}_0$ such that $P$ is the maximum of $F(P|\hat{P}_0,\sigma)$. Analogously, the maximum likelihood estimator returns $\hat{P}_0$ that maximizes the likelihood $\mathcal{L}(P_0)$ (see \citealt{Vaillancourt_2006} for further details). The bottom panel of Fig.~\ref{fig:app:debias_uncertainty} compares the observed $P \, /\, \sigma$ value and the estimated $P_0 \, / \, \sigma$ for the two different estimators (noting here that the normalization by $\sigma$ is to ease calculation and should not necessarily be taken as the signal-to-noise ratio of the observation; see later). Both of these estimators asymptotically approach $P \approx P_0$ such that at $P \, / \, \sigma \gtrsim 3$, $P_0 \approx \sqrt{P^2 - \sigma^2}$. For low values of $P \, / \, \sigma$, both estimators also return a null value, i.e., the observed signal is most likely only noise.

\citet{Simmons_Stewart_1985} \citep[see also][]{Vaillancourt_2006} described an approach to estimate the uncertainty of $P_0$ based on defining confidence intervals for $P_0$. In their approach they chose to minimize the width of the confidence interval; however, this results in asymmetric intervals about $P_0$, which complicates the plotting of polarization line segments based on their perceived signal-to-noise ratio. Here we opt to define a $\delta P_0$ value such that:

\begin{equation}
    \int_{P_0 - \delta P_0}^{P_0 + \delta P_0} p(P_0^{\prime}|P,\sigma)~dP_0^{\prime} = 0.68\,\,.
    \label{eqn:uncertainty}
\end{equation}

\noindent That is, 68\% of the random draws of $P_0^{\prime}$ fall within $P_0 - \delta P_0$ and $P_0 + \delta P_0$, giving a more comparable definition to the usual rms value used for estimates of $\sigma$. This range is shown as the shaded bands in the bottom panel of Fig.~\ref{fig:app:debias_uncertainty}, while the top panel shows the value of $\delta P_0$ relative to the input $\sigma$. At only moderate $P \, / \, \sigma$ values ($\gtrsim 2$), $\delta P_0$ approaches $\sigma$, a commonly used assumption \citep[e.g.,][]{Hull_Plambeck_2015}. As $P \, / \, \sigma$ approaches the $\sqrt{2}$ limit for a non-zero $P_0$, the uncertainties deviate significantly from the measured $\sigma$ due to the asymmetric form of Eqn.~\ref{eq:app:rice}, resulting in a scenario where only upper limits can be estimated. As the inferred $\delta P_0 \approx \sigma$ for a larger range of $P \, / \, \sigma$ values when using the ``most probable'' estimator, we opt to use this estimator for the work presented in this paper.

\end{document}